\documentclass[]{tJOT2e}

\usepackage{epstopdf}

\usepackage[numbers,sort&compress]{natbib}
\bibpunct[, ]{[}{]}{,}{n}{,}{,}
\makeatletter
\def\NAT@def@citea{\def\@citea{\NAT@separator}}
\makeatother

\usepackage{multirow}
\usepackage{amsmath,amssymb}

\usepackage{epstopdf}
\usepackage{color}

\usepackage{tikz,pgfplots}
\usepackage{float}
\usepackage{graphicx}

\begin{document}



\author{{J. H. Marlow\textsuperscript{a}
and W. Brevis\textsuperscript{b} 
and F. C. G. A. Nicolleau\textsuperscript{a}\thanks{Corresponding author. Email: F.Nicolleau@Sheffield.ac.uk} 
}
\affil{
\textsuperscript{a}SFMG, The University of Sheffield, S1 3JD Sheffield, UK 
\textsuperscript{b}School of Engineering, Pontifical Catholic University of Chile, Chile;
}
}

\title{Wake Characterisation of 3-Dimensional Multiscale Porous Obstacles}


\maketitle

\date{\today}

\begin{abstract}

In this research article we study the wake formation behind 3-Dimensional Multi-scale Porous Obstacles (3DMPOs). Particle Imaging Velocimetry (PIV) is used in a non-shallow ($B/h =1.5$) water flume, with measurements carried out across the $x$ - $y$ plane (at $z$ = 130 mm) and with $Re=70,000$ based on the free-stream velocity ($U_{\infty}$). To characterise the downstream wake characteristics of 3DMPOs the obstacles are split into 3 regimes; (1) non-porous, (2) porous with a single internal scale and (3) porous fractals (based on the Sierpinski carpet). The void fraction ($\phi=0.7$) and external dimensions ($D$) of the obstacles remained constant, whilst the internal geometrical parameters such as fractal dimension ($D_f$), lacunarity ($\Lambda$) and succolarity ($\sigma$) were varied. We are able to identify a relationship between these topological parameters characterising the 3DMPOs and the resultant wake characteristics. The fractal dimension ($D_f$) and lacunarity ($\Lambda$) are found to be responsible for the formation of the downstream steady wake region, whilst the succolarity ($\sigma$) affects the position of the detached low velocity recirculation region. The power spectral energy densities (PSDs) of the 3DMPOs are also seen to be affected by the succolarity ($\sigma$) in case (3), and indicate movement away from Kolomogorov's -5/3 power law. 
\bigskip


\bigskip

\end{abstract}

\section*{Nomenclature}

\begin{tabbing}
  XXXXXX \= \kill
  $B$  \> 			 channel width,            mm    
\\
  $D_f$ \>  fractal dimension 
\\ $ D $  \>   obstacle size (width),            mm  
\\ $D_{G} $ \> 	     array diameter,                mm  
\\ $D_{C} $ \> 	     cylinder diameter,              mm  
\\ $f$  \> 			 wake frequency,                 Hz
\\ $f_{s} $ \> 	     sampling frequency,                Hz
\\ $ h$  \>      		 water depth,                 mm
\\ $ H$  \>       		 obstacle height,                mm
\\ $L_{1} $ \> 	     length of jet streams             
\\ $L_{2} $ \>	     length of steady wake
\\ $L_{3} $ \> 	     total length of wake
\\ $N_C $  	\>     number of obstacles in a patch
\\ $Re$  \> Reynolds number
\\ $St$ \>    	     Strouhal number
\\ $t$            \> time, s
\\ $u$ \> 	 streamwise velocity,  m~s$^{-1}$   
\\ $U_{b} $ \> 	 bulk velocity,  m~s$^{-1}$   
\\ $U_{\infty} $ \> 	 freestream velocity  (at $z=h$),  m~s$^{-1}$            
\\ $\overline{U}_{s} $ \>  average velocity of the shear layer,   m~s$^{-1}$
\\ $x$ \> 	 distance in streamwise direction, mm
\\ $y$ \> 	 distance in spanwise direction, mm
\\ 3DMPO  \>          3-Dimensional Multi-scale Porous Obstacle
\end{tabbing}

\section*{Greek Symbols}

\begin{tabbing}
  XXX \= \kill
\\$\overline{\epsilon}$ \>	     energy dissipation rate, m$^{-3}$ s$^{-3}$  
\\ $\phi$     \>    	 void fraction 
\\ $\Lambda$  \>       lacunarity			
\\ $\sigma$  \>        succolarity			
 \end{tabbing}

\section{Introduction}

Examples of fluid flow through 3-Dimensional Multi-scale Porous Obstacles (3DMPOs) can be found in a number of different environments: airflow through large scale urban areas, tidal flow through offshore wind and tidal farms, as well as in nature \citep{Zong-et-Nepf-2012, Sand-Jensen-et-Pedersen-2008, Jarvela-2002, Kouwen-et-al-1981}. 3DMPOs impinging on bulk flows instigate the formation of high energy downstream turbulent structures, causing destruction to the surrounding environment or aid the mixing processes. Therefore, in an effort to harness these structures in human activity it is important to understand and characterise the effect of an obstacle's internal parameters on the downstream flow regime. Understanding how to influence the downstream wake is beneficial in the realms of social and economic health e.g. urban planning to ensure adequate pollution dispersion in large-scale urban areas or improved mixing in industrial processes. The use of 3DMPOs allows the systematic isolation of unique internal parameters responsible for the downstream turbulent structures. 
\\[2ex]
The flow through and around porous obstacles can be particularly complex. It has been common practice to use only one characteristic internal scale as a control parameter, with numerous studies of flow through cylindrical arrays \cite{Chang-et-Constantinescu-2015, Taddei-2016-PhD, Zong-et-Nepf-2012, Takemura-et-Tanaka-2007}, or individual solid square columns  \cite{Wang-et-Zhou-2009}. 
However, in the real world, these arrays do not consist of a single internal scale. Using 3DMPOs will be the next step in our understanding of 3-Dimensional (3D) porous obstacles impinging on bulk flows. The objective of the research presented here is to identify and evaluate the internal parameters responsible for the phenomena identified in the downstream wake in a non-shallow bulk flow. 
\\[2ex]
The paper is organised as follows: in \S\ref{secBT} we present the background theory,  \S\ref{secexp} describes the experimental set up. The results are discussed in \S\ref{secres} and summarised in the conclusion (\S\ref{seccon}).

\section{Background Theory \label{secBT}}

The literature currently available regarding 3D obstacles predominantly focuses on a single internal scale or non-porous obstacles. Nevertheless, there have been numerous studies into characterising these flows which we discuss in this section. The study of solid obstacles in bulk flow is a well-discussed topic, with the effect of the aspect ratio ($h/D$) on the downstream wake being one of the main parameters considered \cite{Sakamoto-et-Arie-1983, Okamoto-Sunabashiri-1959, Wang-et-Zhou-2009}. The span-wise vortices identified in these studies are of particular interest to our study owing to our experimental setup, with the antisymmetric span-wise vortex shedding being responsible for the formation of the well-known von K\'{a}rm\'{a}n vortex street. \citeauthor{Chen-Jirka-1995} \cite{Chen-Jirka-1995} performed a similar experiment studying the shallow 2-dimensional turbulent wake flows using solid cylinders. The experiment was performed in a shallow flow. They identified three regimes of flow around a solid obstacle: i) a vortex street with an oscillating vortex shedding mechanism ii) an unsteady near-wake bubble with wake instabilities growing downstream, and iii) a steady bubble with a steady wake. 

One of the earliest studies into wake formation behind porous obstacles was carried out by \citeauthor{Castro-1971} \cite{Castro-1971}  who identified the porosity threshold as being the parameter responsible for the transition between two flow regimes in 2D plates, with and without vortex shedding.
 
The effect of the void fraction ($\phi$) 
\begin{equation}
    \phi= \frac{\text{volume of voids in the object}}{\text{total volume of the object}}
\label{voidefinition}
\end{equation}
was studied in  \cite{Nicolle-Eames-2011} using high resolution numerical simulations. In these simulations, the obstacles contained a number of equally spaced cylinders ($N_C$), each with the same diameter $D$. \citeauthor{Nicolle-Eames-2011} identified 3 flow regimes dependent on $\phi$, which were: i) at low  void fractions ($\phi<0.05$) individual wakes patterns formed for each individual cylinder ii) at medium void fractions ($0.05<\phi<0.15$), a shear layer formed resulting in a steady force acting on the obstacle iii) at high void fractions ($\phi>0.15$) the patch produced a vortex street, similar to a solid body. 
\\[2ex]
\citeauthor{Chang-et-Constantinescu-2015} \cite{Chang-et-Constantinescu-2015}  also carried out numerical simulations varying the $\phi$ of a cylindrical array using large eddy simulation (LES). Their study, when comparing to \citeauthor{Nicolle-Eames-2011}, also identified that the porosity of the patch is responsible for the extended steady wake region and shear layers. Two flow regimes were identified; for $\phi>$0.05 it predicted low-frequency large-scale wake billows comparable to a von K\'{a}rm\'{a}n street usually observed behind a solid obstacle, and as $\phi$ increased the steady wake and reattachment is shifted downstream. However, for $\phi<$0.05 the interaction between the wakes of each individual solid cylinder was weak and no steady region or separated shear layer (SSL) was formed and as a result. The wake billows were believed to be formed because of the instability within the SSLs causing a phased undulating motion of a large scales that deform the SSL. 
%
%
As can be seen from this brief literature survey, in the majority of arrays studied, it is common to use solid obstacles or single scale cylinder arrays, due to the interest in modelling vegetation and river structures. By contrast, our study uses square porous columns maintaining a constant void fraction whilst varying internal scales. This is of significance as it draws closer parallels to man-made structures such as large-scale urban areas.
\begin{figure}[h]
\centering
\includegraphics[width=0.75\textwidth]{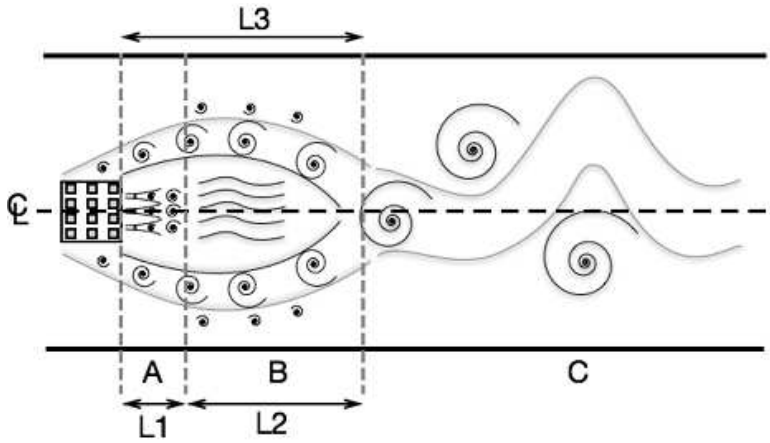}
\caption{Predicted wake behind a porous obstacle.}
\label{wakepatternfig}
\end{figure}
\\[2ex]
From the literature review a detailed picture can been constructed of the predicted wake behind the 3DMPOs. This is
illustrated in Figure~\ref{wakepatternfig}:
\begin{enumerate}
\item
In certain flow regimes ($0.05<\phi<0.15$), the porosity of the obstacles causes the formation of a unique wake immediately downstream of the obstacle. The incoming flow percolates through the object forming a region of low-velocity steady flow downstream of the obstacle, which is known as the steady wake, which would vary in size with $\phi$.  \citeauthor{Zong-et-Nepf-2012} \cite{Zong-et-Nepf-2012} proposed a method to define the length of the low-velocity zone. This was done by estimating the wake based on their plane shear layer ($L_2$) growth equation:
\begin{equation}
    L_{2} \approx \frac{D/2}{S_{\delta}}\frac{\overline{U_s}}{\Delta {U}}
\end{equation}
where $\overline{U_s}$ is the average velocity of the shear layer and 
$S_\delta$ an empirical parameter. 
\item
However, the multi-scale porous nature of the obstacles adds new characteristic scales as jets of fluid are created by the individual scales inside the obstacle as the flow exits generating small-scale turbulence $L_1$, pushing the steady region downstream. The steady wake region dissipates the small scale turbulence creating a constant velocity region $L_2$ in the downstream direction until the formation of a von K\'{a}rm\'{a}n street.
\item 
In regards to the flow around the boundaries of the obstacle, the flow that passes through the obstacle has a lower velocity than the flow that passes around the obstacle. This causes separated shear layers to form around the outside of the obstacle which grow as they progress downstream, but do not interact until the reattachment zone at the end of the steady region ($L_3$). Once the shear layers interact a von K\'arm\'an vortex street is formed from their interaction. The von K\'arm\'an vortex street then grows as it travels downstream until it finally interacts with the walls of the confined flow. 

\end{enumerate}


 \section{Experiment \label{secexp}}
 \subsection{3-Dimensional Multi-Scale Porous Obstacles (3DMPOs)}
 
\begin{figure}[h]
\includegraphics[width=\textwidth]{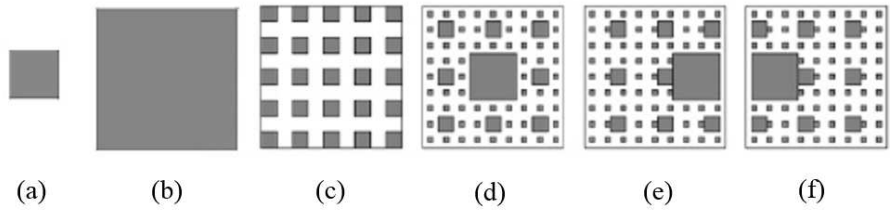}
\caption{Sketch of each of the 3DMPOs used in this study (fluid flow left to right). Case (1) is (a) \& (b) used as the baseline solid obstacles, Case (2) is (c) a single internal scale and Case (3) is ($d$), ($e$) \& ($f$) the Sierpinski carpet based fractals.}
\label{objects-fig}
\end{figure}

In this experiment we use the six obstacles shown in Figure~\ref{objects-fig} a-f to cover a range of internal parameters. PIV is used to measure the instantaneous velocity field generated by each obstacle in the downstream near wake, allowing the identification of significant turbulent structures underlying the flow.

To characterise the internal structure of each obstacle, four complementary parameters were used: void fraction($\phi$, Equation~\ref{voidefinition}), fractal dimension ($D_f$), Lacunarity ($\Lambda$) and Succolarity ($\sigma$). 

 The fractal dimension $D_f$ is defined as a box-counting dimension:
\begin{equation}
   D_f=- \frac{\Delta{\ln{N(r)}}}{\Delta{\ln{(r)}}}
\end{equation}
where $N(r)$ is the number of objects of size $r$ needed to cover the total object. The Lacunarity ($\Lambda$) in the case of the 3DMPOs allows the classification of the size and frequency of the spacing within the fractal obstacle. If a fractal has large intervals, then it is said to have high $\Lambda$. We use the definition from Allain and Cloitre \citep{Allain-et-Cloitre-1991} who proposed the gliding box mechanism for calculating the Lacunarity.
The Succolarity ($\sigma$) is an additional measure used to characterise fractal shapes. It allows classification between different textures that have the same $D_f$ and $\Lambda$ by defining the obstacles ability to allow through-flow, with a higher Succolarity resulting in increased through flow, using the definition from de Melo and Conci \citep{de-Melo-et-Conci-2013}. 
\\[2ex]
Purposefully the four porous obstacles of our series have the same porosity to allow the multi-scale parameters of $D_f$, $\Lambda$ and $\sigma$ to be assessed. These obstacles were of indefinite height, emerging from the flow in $z$, resulting in no overtopping flow and the downstream wake was confined within the flume boundaries. Table~\ref{Obstacle Information} summarises each of the obstacle parameters, it can be seen that the void fraction is large for the porous obstacles ($\Phi$ $\approx 0.7$). These obstacles will lie in the third regime proposed in \cite{Nicolle-Eames-2011}, and it can be expected that the high void fraction will result in the formation of a von K\'{a}rm\'{a}n vortex street downstream of the obstacle. 
\begin{table}[h]
\begin{tabular}{lcccccc}
\hline
    & {Obstacle}     & 
    {\begin{tabular}[c]{@{}c@{}}Void \\ Fraction\\ ($\phi$)\end{tabular}} &
    
     {\begin{tabular}[c]{@{}c@{}}Fractal \\ Dimension\\($D_f$)\end{tabular}} & 
     
     {\begin{tabular}[c]{@{}c@{}}Lacunarity\\($\Lambda)$\end{tabular}} & 
     
     {\begin{tabular}[c]{@{}c@{}}Succolarity\\ Upstream to \\ Downstream\\ ($\sigma$)\end{tabular}} \\ \hline 
(a) & First Iteration & 0 &-&- &-                                                              \\ 
(b) & Solid Column & 0 & -& -&-                                                                                                                        \\ 
(c) & Uniform Obstacle  & 0.69 & - & 0.08 & 0.691                                                                                                                                 \\ 
(d) & Sierpinski Carpet & 0.7 & 1.89 & 0.14 & 0.702                                                                                                                                    \\ 
(e) & Downstream Obstacle & 0.7 & 1.89 & 1.22 & 0.642                                                                                                                                   \\ 
(f) & Upstream Obstacle & 0.7 & 1.89  & 1.22 & 0.763                                                                                                                                                                                                                                                             \\ \hline
     \end{tabular}
\caption{\label{Obstacle Information}Parameters characterising each obstacle.}
\end{table}
\\
\textbf{Obstacle (a)} represents the largest scale present in (d), (e) and (f). This obstacle was purposefully made and included in the experiment to identify its vortex street oscillating frequency and the turbulent scales it is responsible for in the flow. This would then allow a comparison between (a) and the 3DMPOs, (d), (e), and (f). 
\newline
\textbf{Obstacle (b)} is a solid single column of the same size ($D$) as the 3DMPOs. This obstacle was purposefully made to allow comparison between a solid obstacle and the porous obstacles in the flow.
\newline
\textbf{Obstacle (c)} is a uniformly spaced porous obstacle with only one internal scale which can be used for comparison against the multiscale obstacles of (d), (e), and~(f).
\newline
\textbf{Obstacle (d)} is a Sierpinski carpet based obstacle
\cite{Adler-1987}.
\newline
\textbf{Obstacles (e)} and \textbf{(f)} are variations of Sierpinski's carpet with the largest iteration being moved downstream and upstream respectively.

\subsection{Experimental Set-up}

\begin{figure}[h]
\centering
\includegraphics[width=0.9\textwidth]{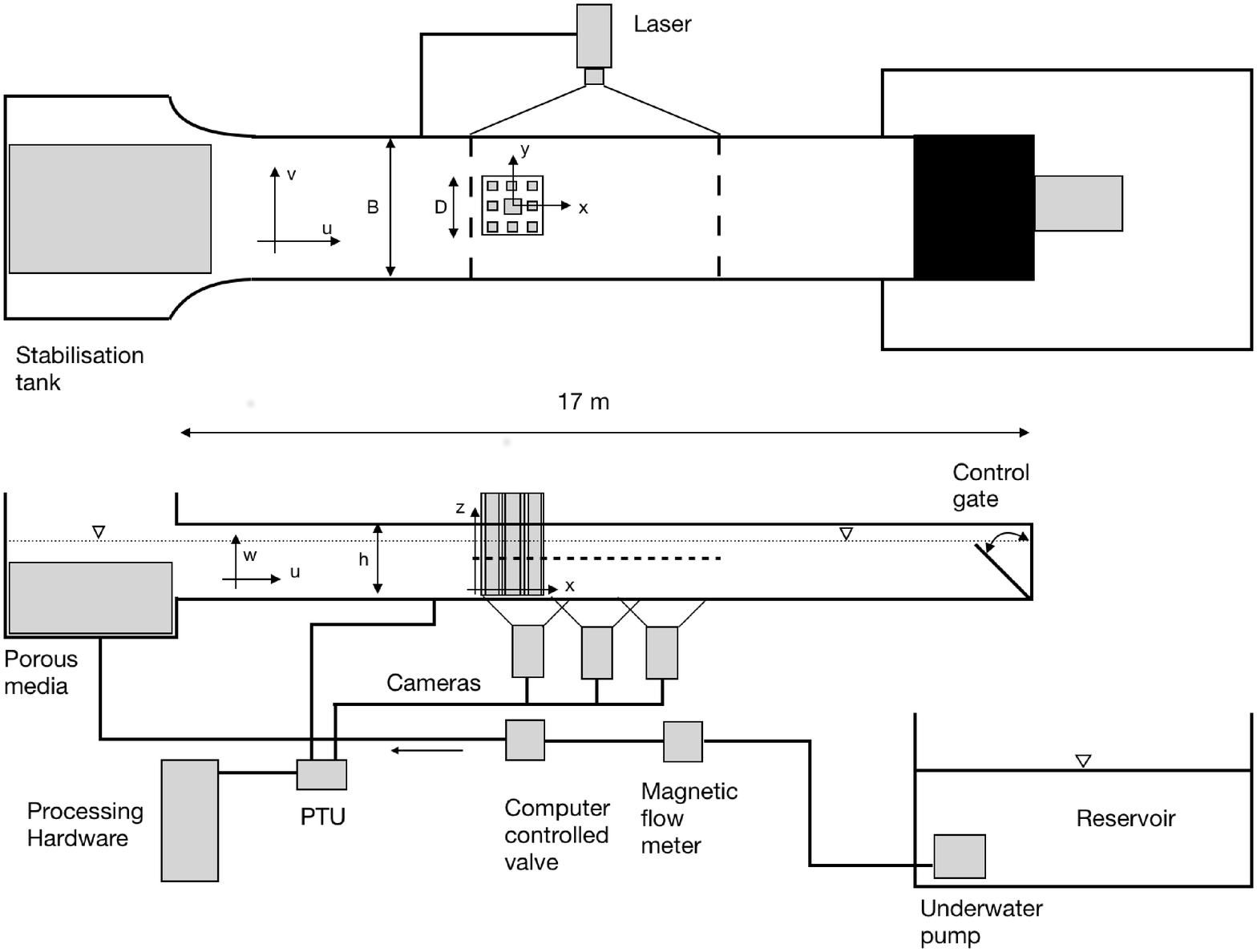}
\caption{Schematic of experimental setup.}
\label{expfigcam}
\end{figure}

The experimental set-up is presented in Figure~\ref{expfigcam}.
The experiment was performed in a closed-loop water flume of cross section 0.486 m $\times$ 0.480 m with a 1 m long test section, under non-shallow conditions ($B/h$ = 1.5). The obstacles were placed on the centreline and mounted to the base of the flume, with the leading edge 5~mm in from of the start of the test section. The obstacle had a width ($D$) of 135~mm and considered of indefinite length in $z$. The experiments were run at a constant flow rate and water depth of {$Q=29 \, \ell$ s$^{-1}$} and $h=326$~mm respectively. The measurements plane was positioned at 0.4$h$ from the base of the flume to give a bulk velocity $U_{b}$ of {0.183}~m~s$^{-1}$ and a Reynolds number $Re={U_\infty} D/\nu$ = 72405, where $\nu$ is the kinematic viscosity of the fluid and $U_\infty$ the freestream velocity. The experiment parameters are summarised in Table~\ref{tabexpsum}.
\begin{table}[h]
\centering
\begin{tabular}{lllll}
\hline\hline
\\ $B$  &     Channel Width(mm)	& 486
\\ $D$  &        Obstacle Width(mm)    &  135
\\ $h$ & 	       Water Depth (mm)  & 326 &
\\  $U_b$ &      Bulk Velocity(m s$^{-1}$)	  &   {0.183}
\\  $U_{\infty}$ &      Freestream velocity (m s$^{-1}$)	  &  0.445
\\ $Q$  &         Flow Rate ($\ell$~s$^{-1}$) 		& {29}
\\ $B/D$ & 	    Channel Ratio        &3.6 &
\\ $B/h$  &        Non-shallow Condition   & 1.5
\\ $h/D$  &        Aspect ratio   & 2.4
\\ $Re$  &       Reynolds Number      & 72405\\
\hline\hline
\end{tabular}
\caption{\label{tabexpsum}Experimental parameters.}
\end{table}

\subsection{Particle Imaging Velocimetry}

The three synchronised cameras (\textit{Imager MX 4MP}) were mounted below the flume
(Figure~\ref{expfigcam}).  The flow was seeded with polyamide-12 spheres 100~$\mu$m diameter and a density of $1.016 \times 10^3$~kg~m$^{-3}$. The illumination was provided by a double pulsed \textit{Nd:YAG} 200~mJ laser. A crystal powered laser set at 44\% power for Lasers 1 and 2, an attenuation of 58\% and a firing rate of 100~Hz. 
The PIV measurements were taken on a horizontal plane ($x,y$) at  $z$ = 130 mm, with a horizontal laser plane size of 600 $\times$ 900~mm. 
A particle concentration of at least five particles in a region of $32 \times 32$ pixels was enforced to allow effective correlation of the particles, whilst allowing the particles to represent the turbulent structures within the flow \cite{Adrian-et-Westerweel-2011}. 
To ensure effective pre-processing, the particles were required to move approximately 5 pixels per image. With a bulk flow velocity $U_{b}$ = 0.183 m s$^{-1}$ using a sampling frequency $f_s$ = 100 Hz leading to a linear displacement of approximately 8~pixels per snapshot.
The acquisition rate was set to 100 Hz and for a time 72 s to obtain 7200 images. 
\\[2ex]
The correlation algorithm implemented was the direct Fast Fourier Transform correlation (\textit{FFT window deformation}) method. This correlation function allowed multiple passes and dealt well with deformed windows, whilst saving the CPU memory by calculating the correlation iteratively. 
The cameras' area of vision overlapped to ensure the whole flow was captured.
A calibration image was recorded to be used to convert pixel position to real world spatial position and compensate for image distortion.  A conversion algorithm was used to convert the pixel displacement output 
to spatial displacement. 
To filter the data, the Proper Orthogonal Decomposition Detection and Estimation Method (PODDEM) \cite{Higham-et-al-2016} was implemented. 
%

\subsection{Turbulence Statistics}
To understand and visualise the complex turbulent flow, turbulent statistics were based on the Reynolds decomposition:
\begin{equation}
   U=\overline{u}+u'
\label{Reydeceq}
\end{equation}
breaking down the flow into its mean $\overline{u}$ and a fluctuating component $u'$. The time average of the flow was taken to obtain the mean flow field,
\begin{equation}
\langle{U(x_1,y_1,z_1)}\rangle_t=\frac{1}{T}      \sum_{i=1}^{N}U(t_i;x_1,y_1,z_1)\Delta{t} 
\end{equation}
From this we can then define the Reynolds Shear Stress (RSS) tensor
$R_{ij}=\overline{{u_i' u_j'}}$
where $u_i'$, $u_j'$ are components of the velocity fluctuation.
$R_{ij}$ can be used to identify shear layers in the flow, which was extremely useful in this research, owing to the expected presence of the two separated shear layers. The turbulent kinetic energy $K$ can also be calculated from these fluctuating velocity components identified from the Reynolds decomposition. 
\begin{equation}
K = \frac{1}{2} \left ( \overline{{u'}^2}+\overline{{v'}^2}+\overline{{w'}^2} \right )
\end{equation}
\\[2ex]
The energy spectrum $E(f)$ is computed to measure the distribution of energy in the the range of frequencies present in the turbulent flow. It is a useful tool to assess the turbulence characteristics as it can be compared to the well-known Kolmogorov power law, 
\begin{equation}
  E(f)=C_K\overline{\epsilon}^{2/3}f^{-5/3}
\end{equation}
where $C_K$ is Kolmogorov's constant, $f$ is the frequency, $\overline{\epsilon}$ is the dissipation rate. 
%
The  Kolmogorov -5/3 power law can then be used as a bench mark. The largest peak in the power spectrum represents the vortex shedding frequency $f_{peak}$ of the turbulent flow, which allows the calculation of the Strouhal number $St$:
\begin{equation}
  St=\frac{f_{peak}L}{U}
\label{Stouhaldefeq}
\end{equation}
$St$ is a dimensionless number characterising the oscillating flow mechanisms associated with vortex shedding. It measures the unsteadiness of the flow, as a ratio between inertial forces caused by the unsteadiness and inertial forces related to the steady flow. 
Vortex shedding occurs behind a bluff body that is impinging a steady flow, this alternating vortex shedding leads to an oscillating flow structure forming behind the bluff body known as a von K\'{a}rm\'{a}n street. The von K\'{a}rm\'{a}n vortex street is formed through the interaction of the two separate shears layers that form as the flow passes around the bluff body.

\section{Results \label{secres}}

The results presented in this section were obtained from the data collected using the PIV system described in section 3.3. Two runs for each obstacle were carried out to ensure reliable results and good quality images.
The results have been non-dimensionalised as follows:
\\
the lengths have been non-dimensionalised using the obstacle overall characteristic size, 
\begin{equation}
x^* = x/D, \,\,\, y^* = y/D;
\end{equation}
the velocity associated to the freestream flow is used to non-dimensionalise velocities and correlations,
\begin{equation}
u^*=\overline{u}/U_{\infty}, \,\,\, v^* =\overline{v}/U_{\infty}
\end{equation}
\begin{equation}
R_{ij}^* = \overline{{u_i' u_j'}}/{U_{\infty}}^2
\end{equation}
and 
\begin{equation}
K^* = \frac{1}{2} \frac{\overline{{u'}^2}+\overline{{v'}^2}+\overline{{w'}^2}}{{U_{\infty}}^2}
\end{equation}

\subsection{Freestream Characterisation}

\begin{figure}[H]
\begin{minipage}{1\textwidth}
\centering
        \input{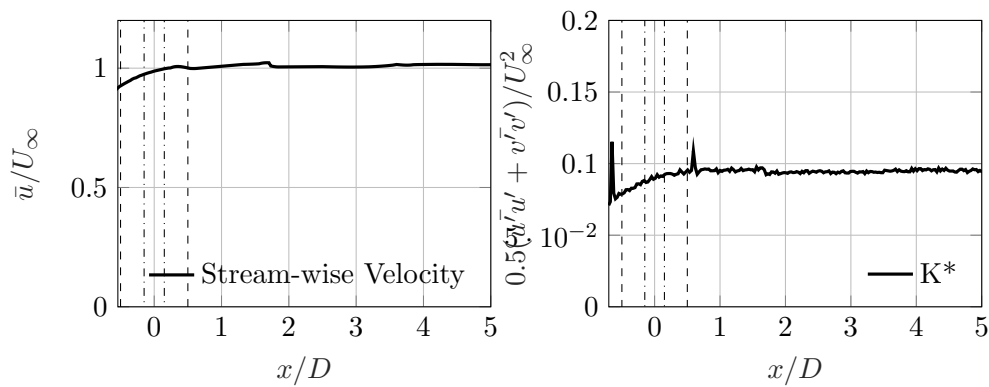}
\end{minipage}
\\
\caption{Characterising the freestream: left mean stream-wise velocity along the centreline of the fume; right non-dimensionalised TKE.}
\label{freestreamfig}
\end{figure}
    
As can be seen in Figure~\ref{freestreamfig}, the flow is fully developed and remains relatively constant throughout the test section. We can see that there is a two trip feature around the flume mounting system which is responsible for the spikes at +/-0.6.

\subsection{Streamwise Velocity}

\begin{figure}[h]
    \begin{subfigure}[First Iteration]{
       \includegraphics[width=0.48\textwidth]{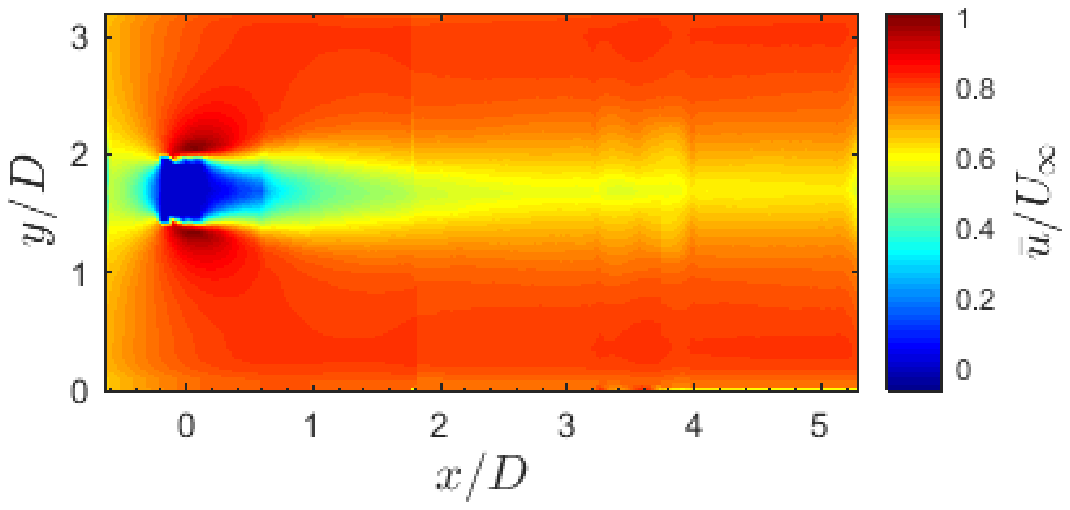}
        }
    \end{subfigure}
     \begin{subfigure}[Solid Column]{
        \includegraphics[width=0.48\textwidth]{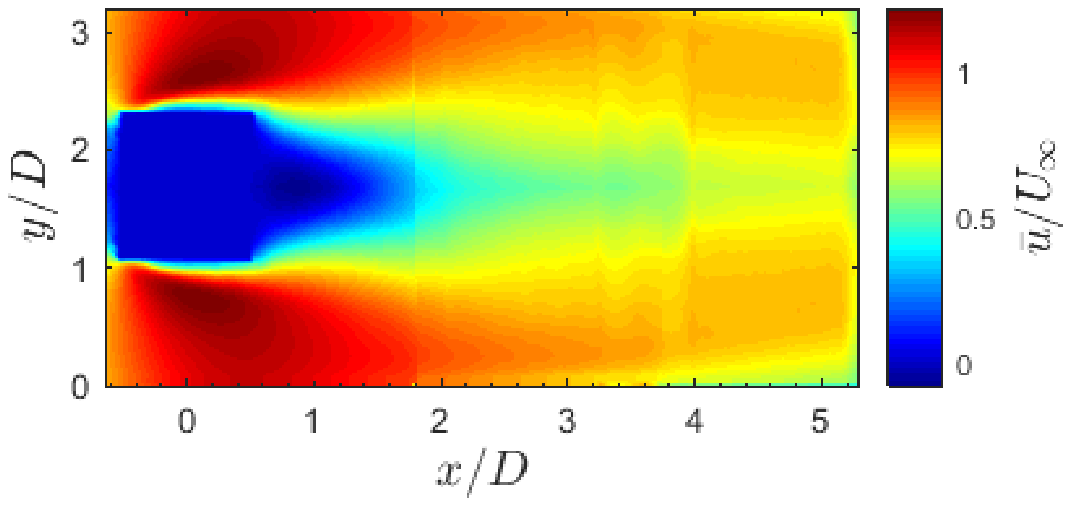}
        }
    \end{subfigure}    
    \begin{subfigure}[Uniform Obstacle]{
        \includegraphics[width=0.48\textwidth]{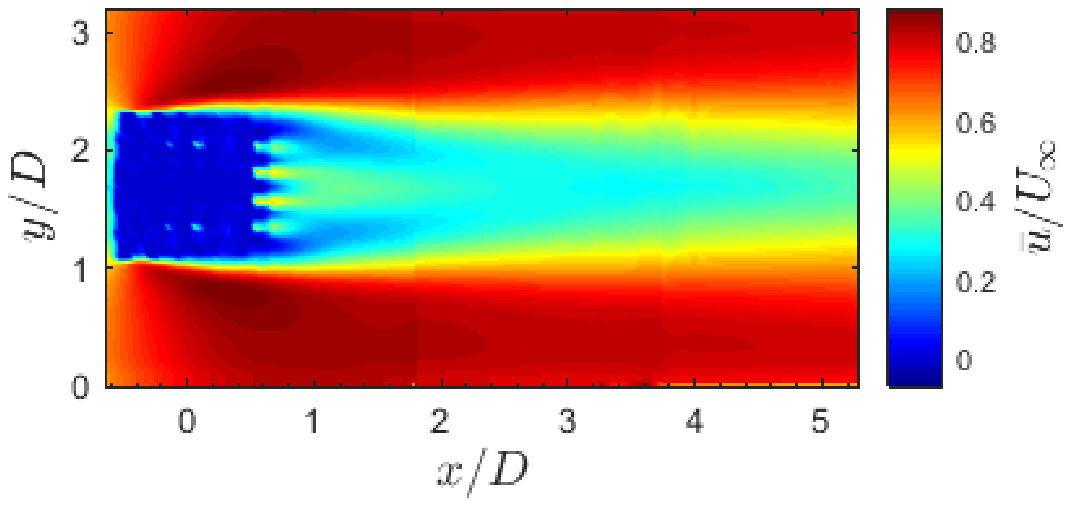}
        }
    \end{subfigure}
    \begin{subfigure}[Sierpinski Obstacle]{
        \includegraphics[width=0.48\textwidth]{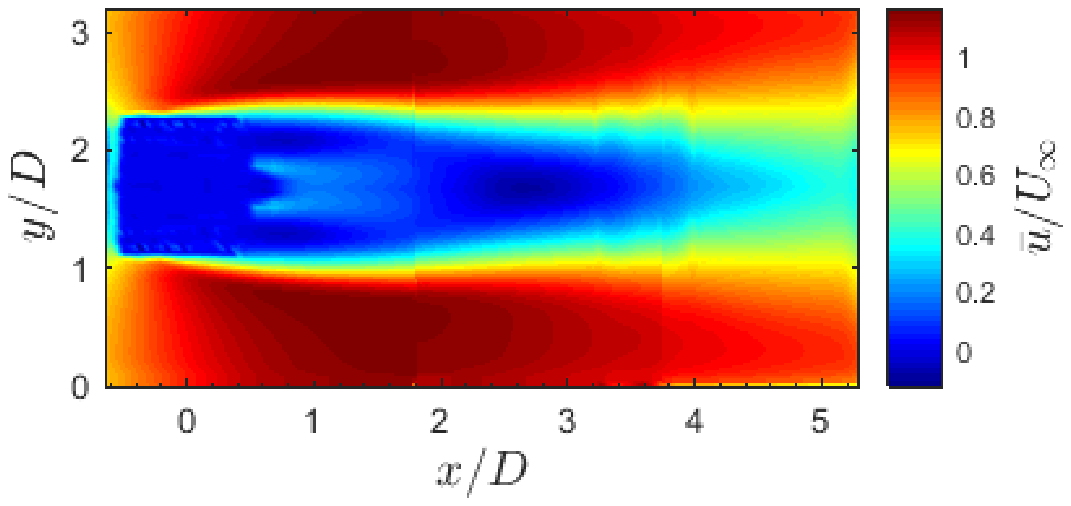}
        }
    \end{subfigure}   
    \begin{subfigure}[Upstream Obstacle]{
        \includegraphics[width=0.48\textwidth]{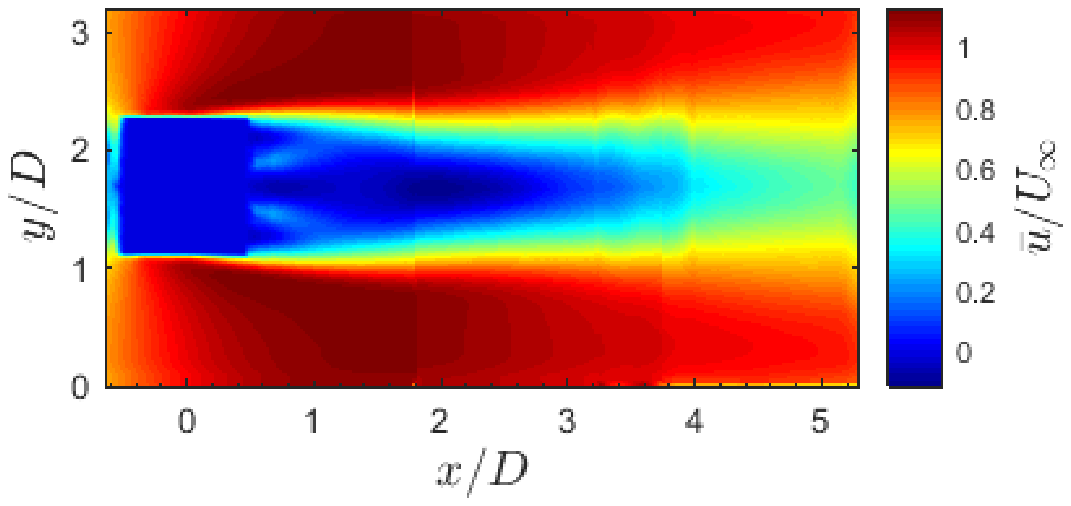}
        }
    \end{subfigure}
    \begin{subfigure}[Downstream Obstacle]{
        \includegraphics[width=0.48\textwidth]{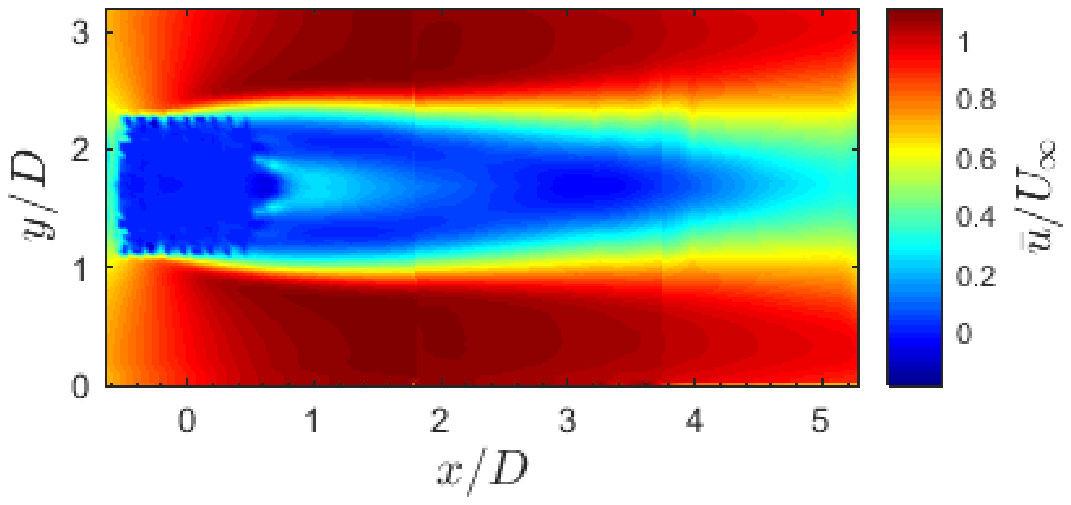}
        }
    \end{subfigure}
     \caption{Streamwise velocity of the wake associated with each obstacle.} 
     \label{Streamvelwakefig}
\end{figure}

Figure ~\ref{Streamvelwakefig} shows the mean streamwise velocity for each of the obstacles.
The solid obstacles ($a$) \& ($b$) were used for comparison owing to the extensive literature on the subject. It can be seen that as expected for the non-porous obstacles (a \& b) a small near wake forms directly behind the obstacle that quickly approaches freestream velocity. This wake contains a small recirculation region of high intensity, that extends $1x^*$ downstream for the solid column ($b$) and $0.5x^*$ for the first iteration (a). As expected the porous obstacles display a completely different behaviour through the formation of an extended steady near-wake region caused by streamwise jet streams leaving the obstacle. Each of the porous obstacles displays its own characteristic wake determined by its internal structure.

The uniform obstacle (c) has a very low Lacunarity ($\Lambda$ = 0.08) representing the large spacing between the internal scales. These large open corridors allow the continuation of the mean flow through the obstacle with only a slight reduction in flow velocity, these jets dominate the recirculation region that would otherwise form behind a solid obstacle. This recirculation region ($L_2$) is broken up into significantly smaller regions of low-pressure zones behind the small scale internal structures present. As a result, the steady near wake structure forming behind the obstacle is of much higher velocity and extends downstream and out of the near field. The internal mid velocity core is surrounded by two low-velocity legs extending from the sides as a result of lateral bleeding. The lateral bleeding is caused by some of the through flow moving laterally into the spaces between the scales. This causes a reduction in flow velocity and this low velocity flow bleeds laterally out into the surrounding flow at the obstacle boundaries forming low-velocity symmetrical legs. 

The increased $\Lambda$ for the Sierpinski based obstacles (d-f) result in a smaller steady wake region. The large $\Lambda$ represents the numerous internal scales present, increasing the resistance to the flow. This results in a reduction of the exit velocity and size of the jet streams. In combination with the largest iteration in the Sierpinski based obstacles, these form a small recirculation region in the low pressure zone behind the large iteration. The position ($L_1$) of the large recirculation zone created by the length scale of the obstacle ($D$) itself is pushed downstream extending, as in the case of the uniform obstacle, by the streamwise jets. This leads to the formation of a mid velocity steady core surrounded by low velocity legs. As the wake extends further downstream the steady core velocity reduces and the low velocity legs creep towards the centreline eventually joining the recirculation region that has been pushed downstream.
\\[2ex]

\begin{figure}[H]
   \input{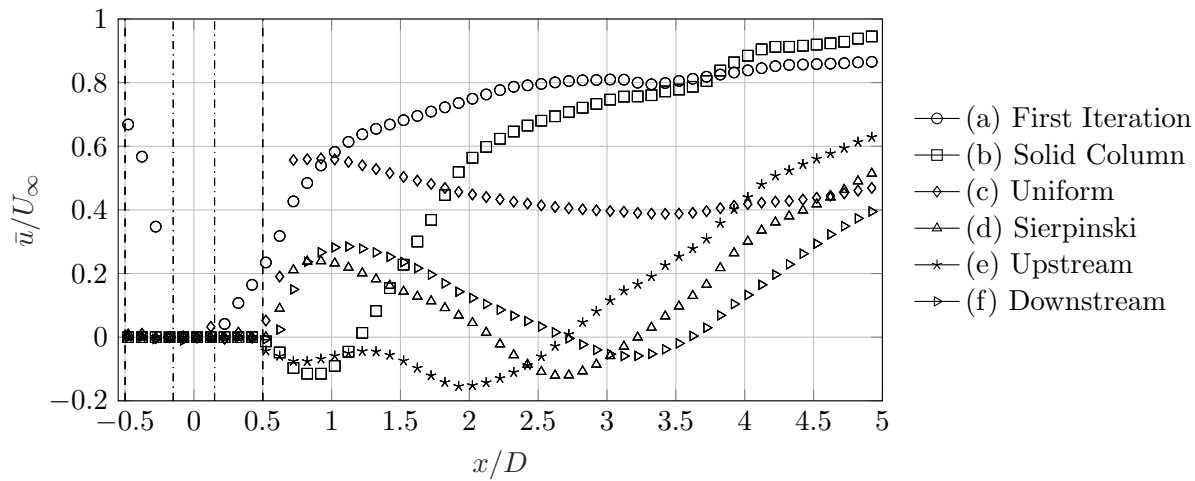}
\caption{\label{umeancentlinefig}Mean streamwise velocity along the centreline for each obstacle.}
\end{figure}

It can be seen from Figure~\ref{Streamvelwakefig}c-f that $L_1$ and resulting position of the detached recirculation region for the Sierpinski based obstacles are dependent on the position of the largest iteration. The recirculation region is indicated by the formation of a ``bowl" of low velocity (see Figure~\ref{umeancentlinefig}) beginning at $2.3x^*$ for obstacle $d$. When the largest iteration is moved downstream ($f$) $L_1$ increases to $2.9x^*$. The largest iteration also seems to create its own recirculation region in the near wake. For the obstacles $e$ and $d$ there is a small recirculation region extending $0.2x^*$ from the obstacle itself. However, for the downstream obstacle, the resultant recirculation region from the largest iteration extends downstream from the obstacle to $2x^*$, joining the two. This can be seen more clearly in Figure~\ref{Streamvelwakefig}d-f. The length of the jets ($L_1$) and low pressure zones ($L_2$) can be seen clearly in the streamwise profiles in Figure~\ref{umeancentlinefig}.
\\[2ex]

\begin{figure}[H]
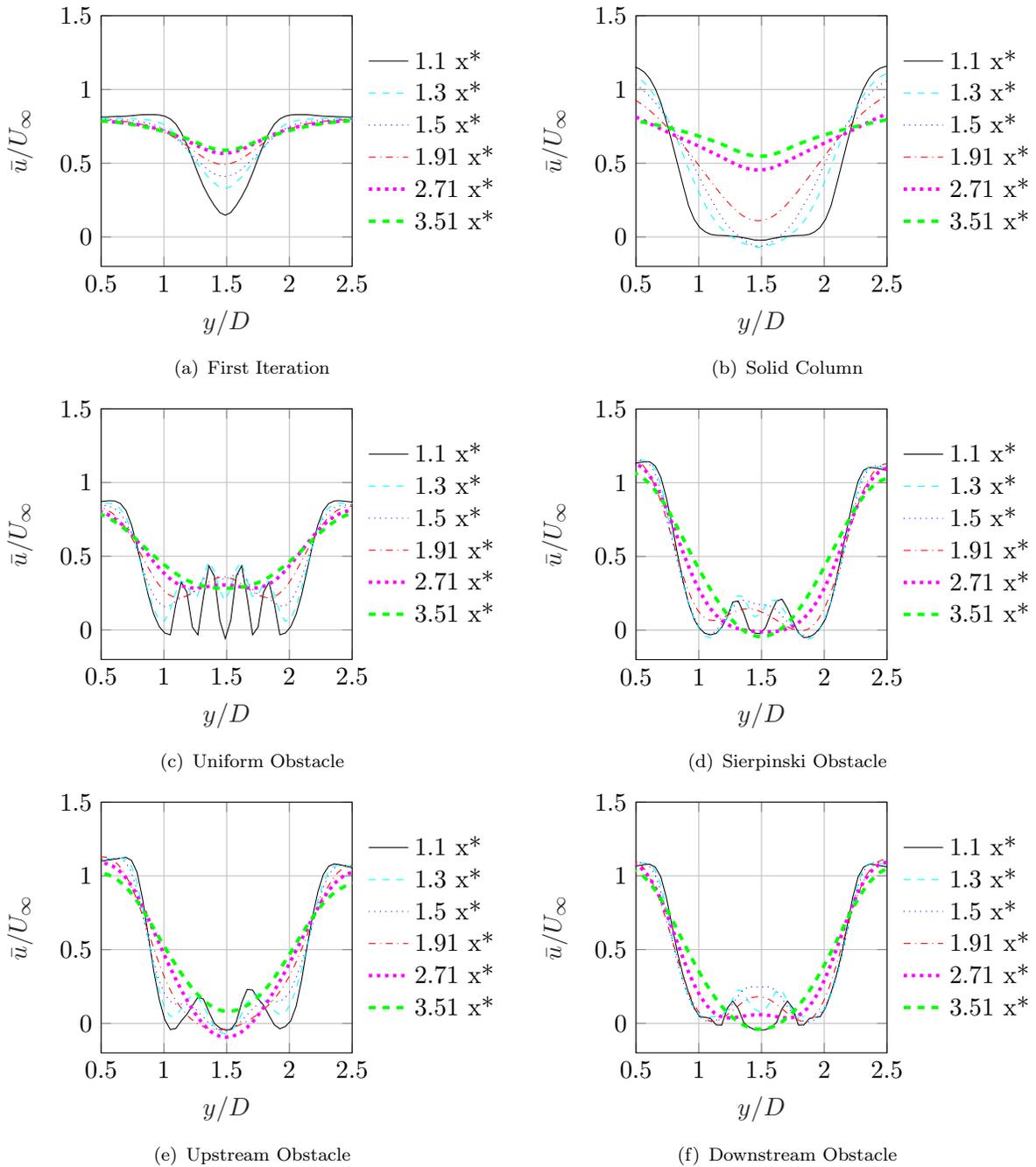

\begin{subfigure}[First Iteration]{
        \input{ulat_FI}
        }
\end{subfigure}
\begin{subfigure}[Solid Column]{
        \input{ulat_SC}
        }
\end{subfigure}
\begin{subfigure}[Uniform Obstacle]{
        \input{ulat_U}
        }
\end{subfigure}
\begin{subfigure}[Sierpinski Obstacle]{
        \input{ulat_SI}
        }
\end{subfigure}
\begin{subfigure}[Upstream Obstacle]{
        \input{ulat_FIU}
        }
\end{subfigure}
\begin{subfigure}[Downstream Obstacle]{
 \input{ulat_FID}

        }
\end{subfigure}
\caption{\label{Umeanspanfig}Mean streamwise velocity profile along $y$ of the downstream wake associated with each obstacle.}
\end{figure}

If we now look to the lateral profiles of the streamwise velocity (Figure~\ref{Umeanspanfig}) the solid obstacles display similar behaviour, with an initial low velocity flow that increases in magnitude and reduces in thickness as it increases in $x^*$ and approaches freestream. It is interesting to see that as the flow moves downstream for the solid obstacle the surrounding freestream velocity also reduces, indicating energy loss that may manifest itself as the drag being applied on the obstacle. The effect of the low $\Lambda$  can be seen in Figure~\ref{Umeanspanfig}c with large jets leaving the uniform obstacle ($c$) as sharp peaks protruding into the higher velocity zone of the figure, which continues until approximately 1.1$x*$ downstream. The low speed legs can also be seen in this profile which are represented by the outer valleys between $1.3x^*$ to $1.91x^*$ downstream, which completely disappear at $3.51x^*$. 

The Sierpinski based obstacles ($d-f$) all display similar profiles in Figure~\ref{Umeanspanfig}d-f. They contain three low velocity zones directly downstream. The central low velocity zone is due to the largest iteration in the obstacle creating its own downstream low pressure zone. The two external low velocity regions are caused by the lateral bleeding as has been also identified in the uniform obstacle. Another interesting phenomenon that is displayed is the appearance of the downstream recirculation zone. For the Sierpinski obstacle ($d$) it can be seen that from $1.5x^*$ the central portion of the velocity profile is increasing but as we move further downstream ($1.91x^*$, $2.71x^*$, $3.51x^*$) this central zone actually decreases in velocity indicating that the profile is entering the detached recirculation region, with a similar pattern being showed by the downstream ($f$) and upstream ($e$) obstacles.

\subsection{Reynolds Shear Stress ($R_{uv}^*$) }

\begin{figure}[htbp]
\begin{subfigure}[First Iteration]{
        \includegraphics[width=0.48\textwidth]{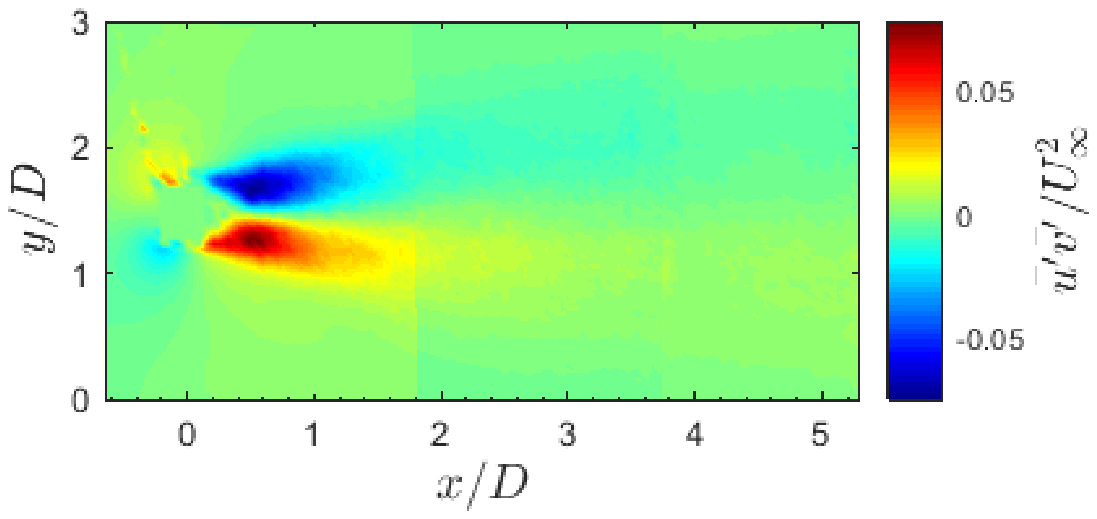}
}
\end{subfigure}
\begin{subfigure}[Solid Column]{
        \includegraphics[width=0.48\textwidth]{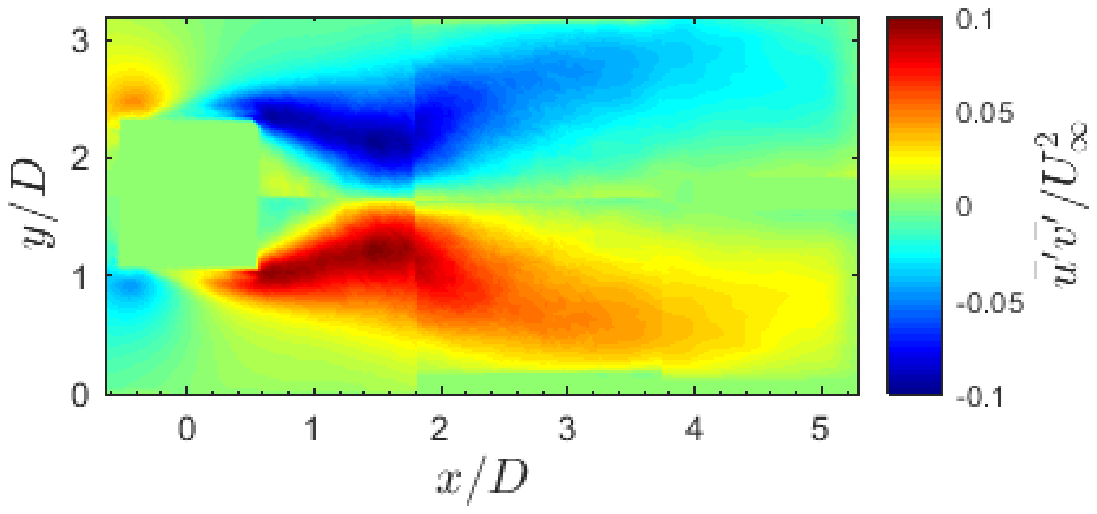}
}
\end{subfigure}
\begin{subfigure}[Uniform Obstacle]{
        \includegraphics[width=0.48\textwidth]{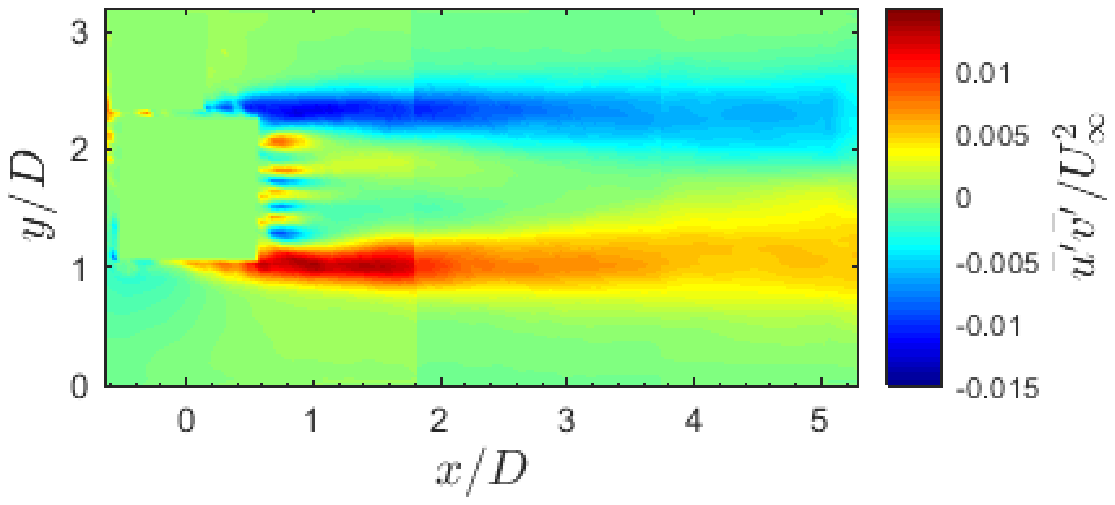}
}
\end{subfigure}
\begin{subfigure}[Sierpinski Obstacle]{
        \includegraphics[width=0.48\textwidth]{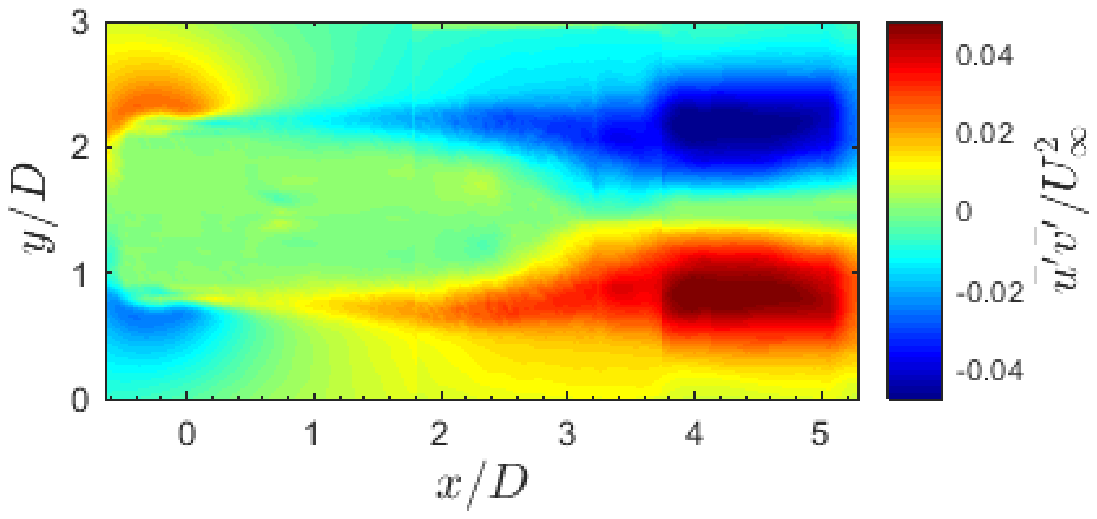}
}
\end{subfigure}
\begin{subfigure}[Upstream Obstacle]{
        \includegraphics[width=0.48\textwidth]{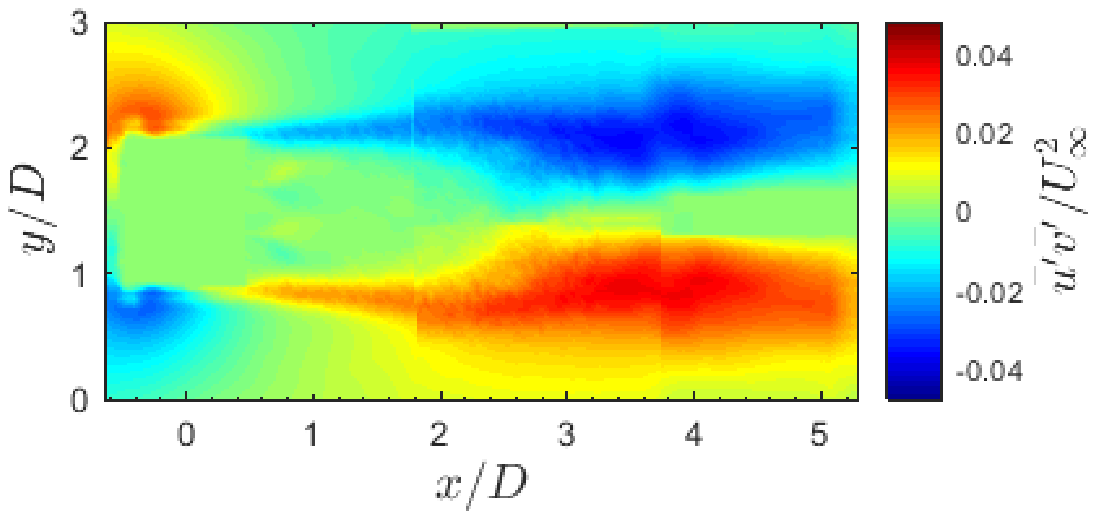}
}
\end{subfigure}
\begin{subfigure}[Downstream Obstacle]{ 
        \includegraphics[width=0.48\textwidth]{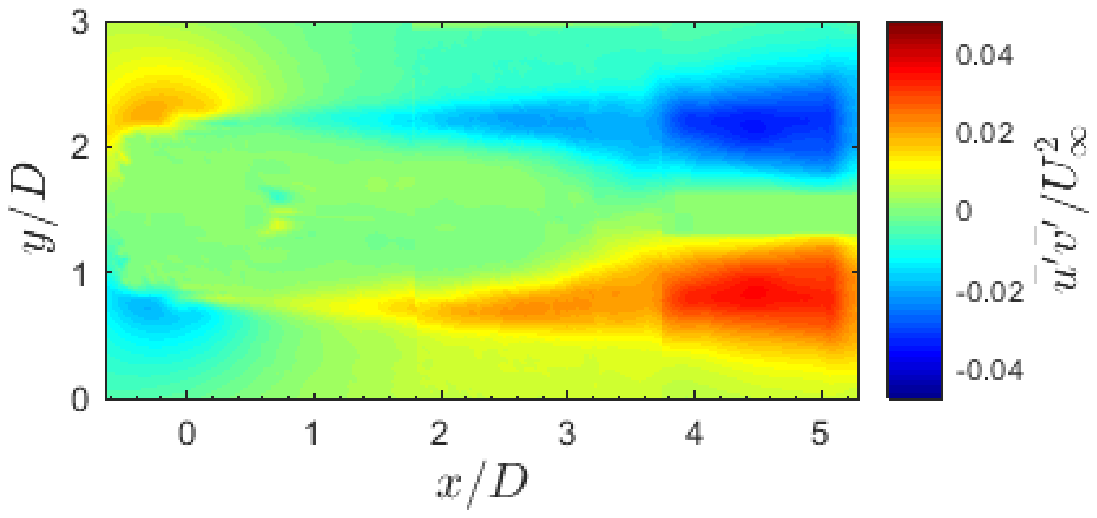}
}
\end{subfigure}
\caption{\label{fig-RST}Normalised Reynold's shear stress $R_{uv}^*$ of the wake associated with each obstacle. 
}
\end{figure}

Figure~\ref{fig-RST} shows iso-value for the normalised Reynolds shear stress $R_{uv}^*=\overline{u'v'}/U_b^2$
based on the velocity horizontal components. 
This normalised Reynolds shear stress allows the identification of the shape and intensity of the shear layers due to the increased stress tensor. The solid obstacles
(a) \& (b) form relatively thick intense separated shear layers that quickly reattach downstream. The maximum value of the shear stress was much greater for the solid obstacles ($\max R_{uv}^*(a)=0.1$ \& $\max R_{uv}^*(b)=0.12$) when compared to the uniform obstacle which displayed a maximum value $R_{uv}^*= 0.015$. 

The uniform obstacle forms much thinner separated shear layers (SSLs) than the other obstacles tested, with the highest intensity much closer to the obstacle itself. The average intensity associated with the shear stress was much lower suggesting small billowing at the edges of the steady wake into the freestream. These SSLs extend a great distance downstream and leave the near field without reattaching. This implies minimal energy transfer in this turbulent behaviour allowing the maintenance of this small scale billowing downstream. 

The uniform obstacle ($c$) also displayed an increase in the shear stress on the streamwise jets leaving the obstacle, which can be seen in Figures~\ref{fig-RST}c at around $x^*=0.8$, which may indicate the presence of small turbulent structures as the jets leave the obstacle. 

The Sierpinski obstacles all displayed a similar SSL thickness of $1y*$ and maximun shear stress intensity $R_{uv}^*=0.05$, but with varying reattachment positions. The SSL thickness of these Sierpinski based obstacles can be compared to the solid column with a thickness of approximately $1y^*$ indicating that is driven by $D$. However, the magnitude of these shear layers were around 50\% of the solid column, as a result of the influence of $\phi$. The formation position of these shear layers and reattachment also varied as discussed in the streamwise velocity section.

\begin{figure}[htbp]
\input{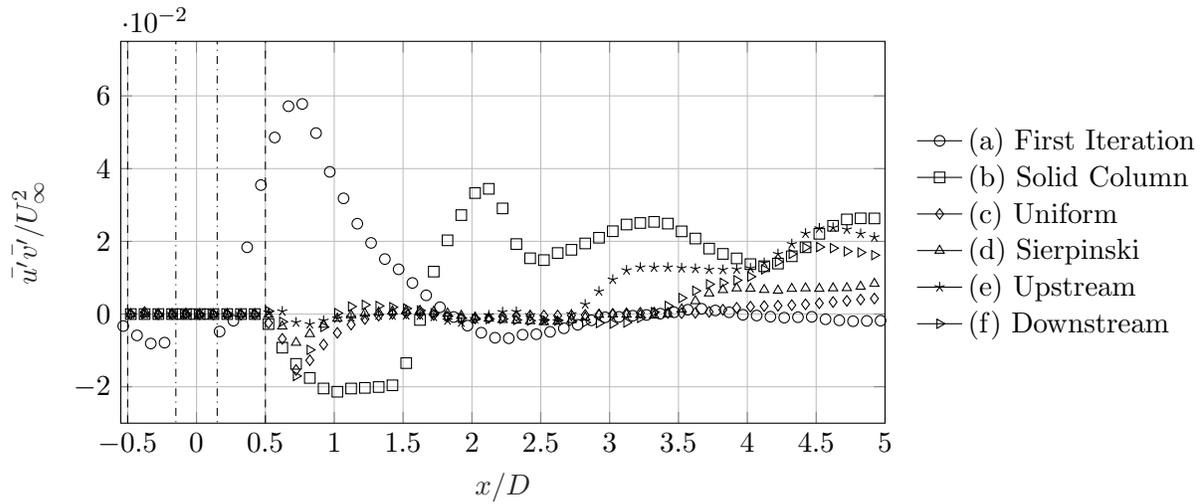}
\caption{\label{fig-RDTCL}Reynolds's shear stress $R_{uv}^*$ along the centreline for each obstacle.}
\end{figure}

Comparison of $R_{uv}^*$ along the centerline is shown in Figure~\ref{fig-RDTCL} for the different obstacles.
Figure~\ref{fig-RDTCL} allows an approximation of the reattachment using the initial increase in $R_{uv}^*$ to indicate reattachment. The uniform obstacle does not reattach resulting in negligible shear stress along the centreline. The solid column ($d$) and first iteration ($a$) reattach quickly at $x^*=1.5$ and $x^*=0.3$ respectively. The Sierpinski obstacles steady wake region extends downstream, with the upstream obstacle ($e$) reattaching first at $x^*=2.7$, and then the Sierpinski ($d$) at $x^*=3.5$ and the downstream obstacle ($f$) at $x^*=3.6$, indicating direct proportionality between iteration position and reattachment. 

\subsection{Turbulent Kinetic Energy of Wake}

\begin{figure}[h]
\input{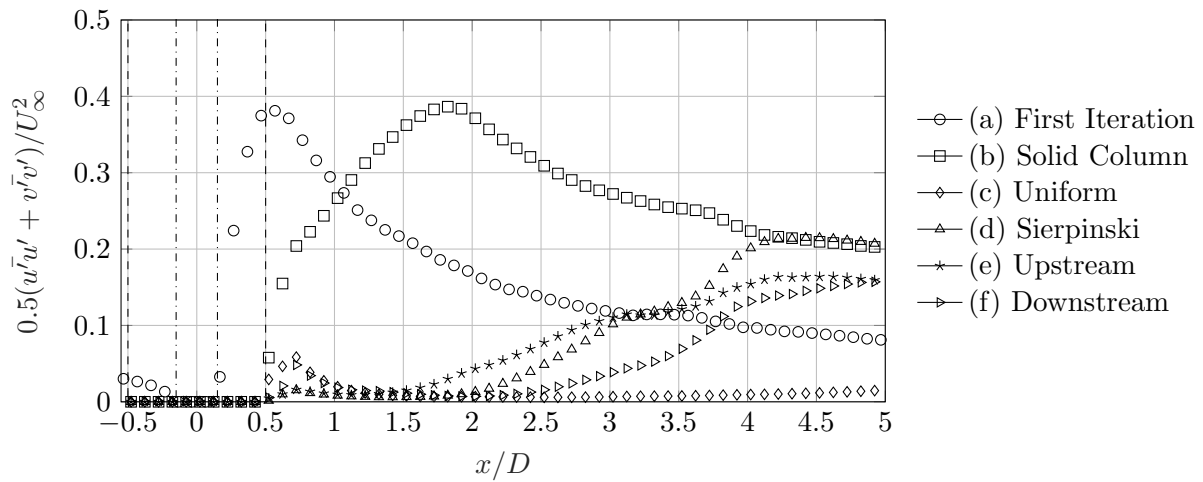}
\caption{\label{TKECLfig}Turbulent kinetic energy $K^*$ along the centreline for each obstacle.}
\end{figure}

As discussed previously, $R_{uv}$ is a useful tool in estimating reattachment and the start of the vortex street, this can be reinforced by analysis of the normalised turbulent kinetic energy $K^*$. Figure~\ref{TKECLfig} shows the development of $K^*$ along the centreline of the wake behind the obstacles. It can be seen in the steady wake region there is negligible $K^*$, however as the flow reattaches and the vortex shedding begins in the von Karman vortex street $K^*$ begins to increase. It can be seen once again that the $K^*$ of the solid obstacles (first iteration ($a$) $\&$ solid column ($b$)) is 2 times greater respectively, than the $K^*$ involved in the vortex shedding behind the porous obstacles. It should also be noticed that the value of $K^*$ associated with the uniform obstacle ($c$) along the centreline is dramatically smaller than that of the other obstacles tested, not rising above $K^*=0.01$ owing to the flow not reattaching within frame.

\begin{figure}[htbp]
\begin{subfigure}[First Iteration]{
        \includegraphics[width=0.48\textwidth]{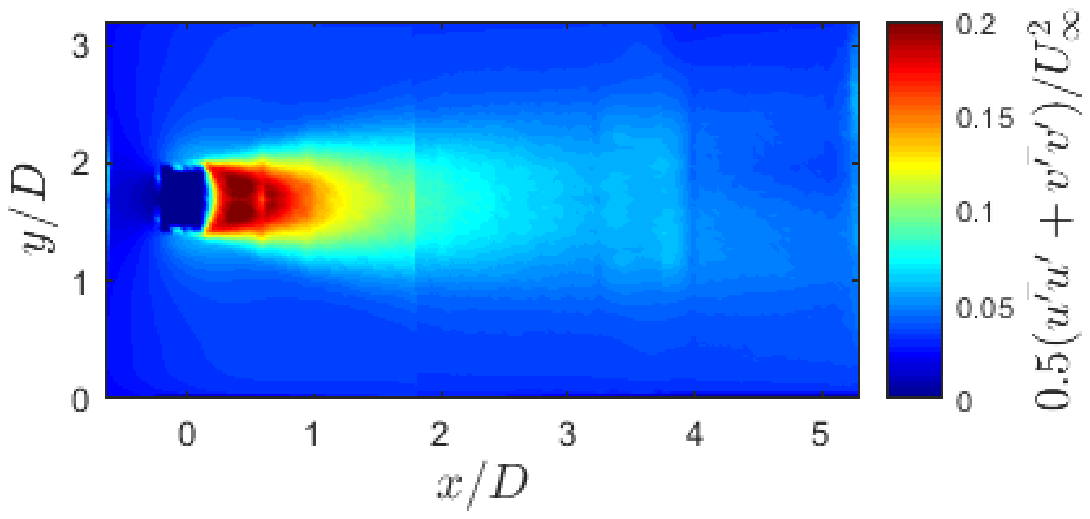}
}
\end{subfigure}
\begin{subfigure}[Solid Column]{
        \includegraphics[width=0.48\textwidth]{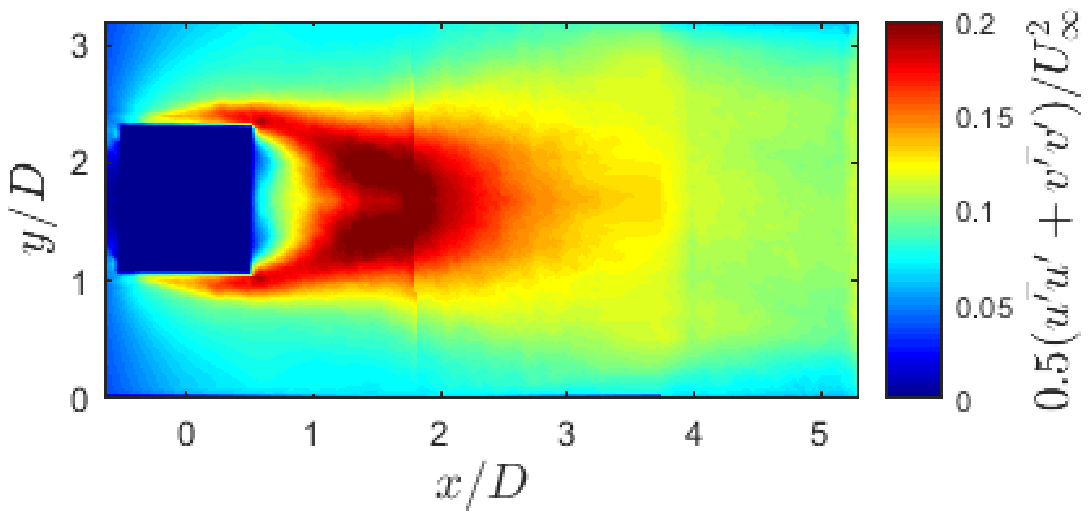}
}
\end{subfigure}
\begin{subfigure}[Uniform Obstacle]{
        \includegraphics[width=0.48\textwidth]{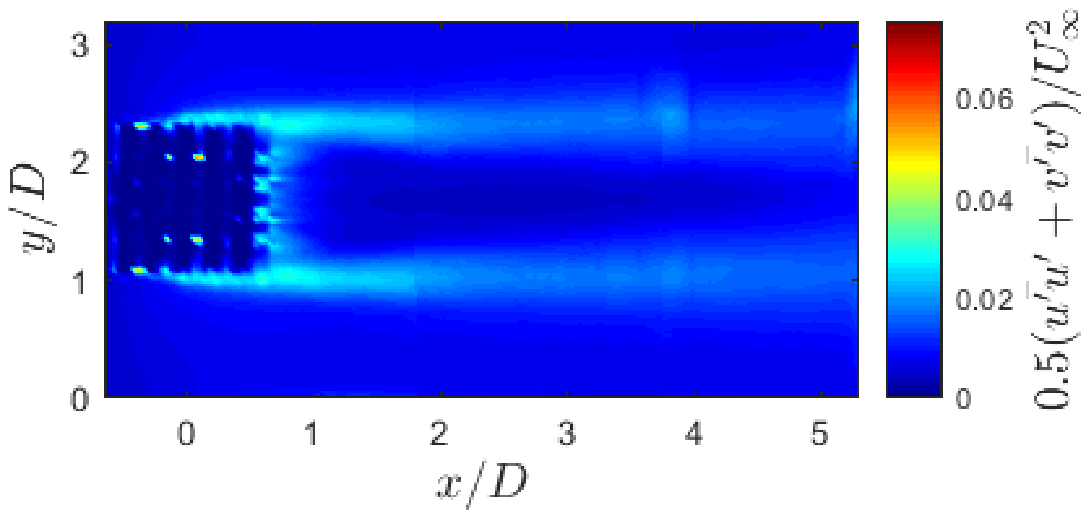}
}
\end{subfigure}
\begin{subfigure}[Sierpinski Obstacle]{
        \includegraphics[width=0.48\textwidth]{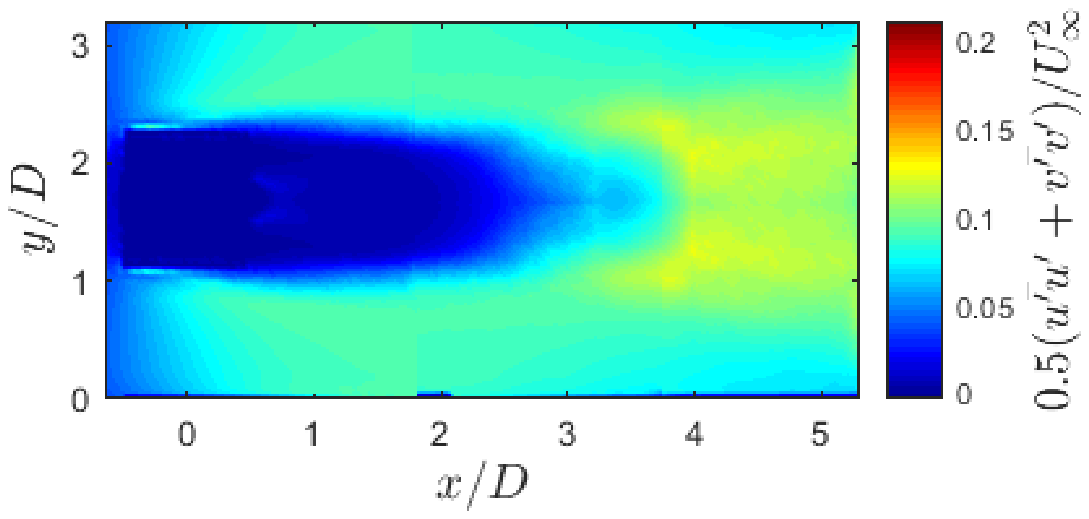}
}
\end{subfigure}
\begin{subfigure}[Upstream Obstacle]{
        \includegraphics[width=0.48\textwidth]{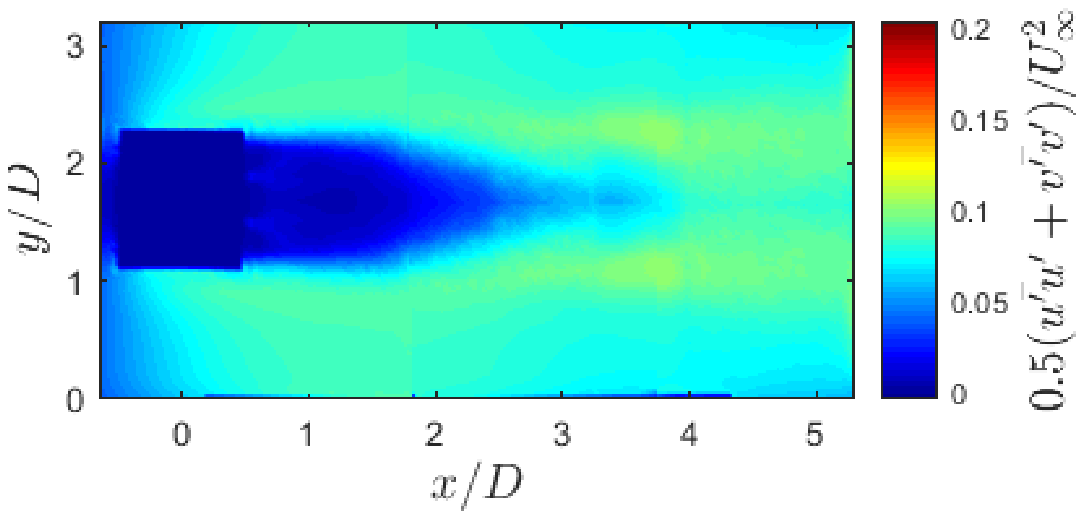}
}
\end{subfigure}
\begin{subfigure}[Downstream Obstacle]{ 
        \includegraphics[width=0.48\textwidth]{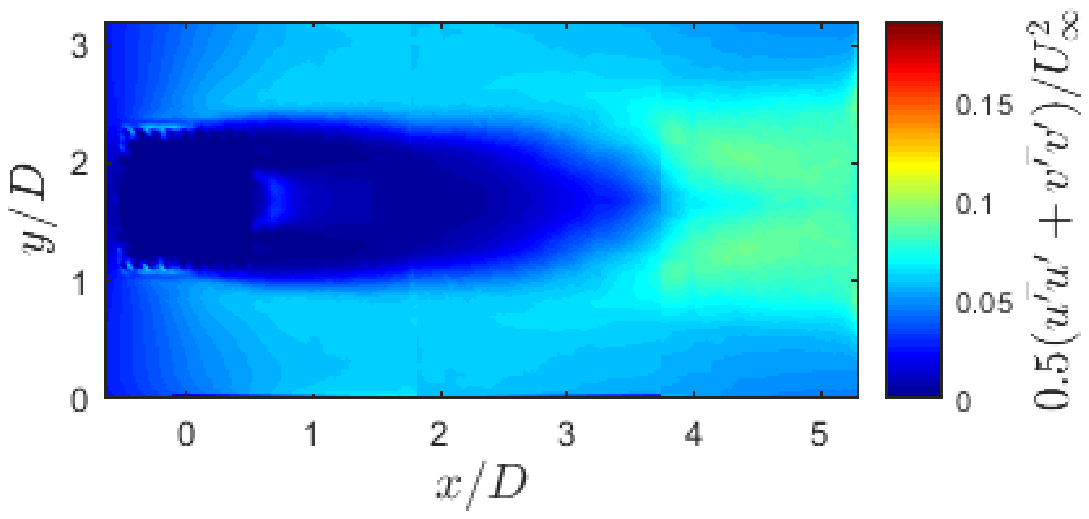}
}
\end{subfigure}
\caption{\label{TKEfig}Turbulent kinetic energy of the wake associated with each obstacle.}
\end{figure}
Figure~\ref{TKEfig} displays the $K^*$-isovalue pattern associated with each of the obstacles tested. 
A region of high $K^*$ represents the energy in turbulent eddies forming in the flow. It can be seen that once again for the uniform ($c$) obstacle, the two turbulent shear layers surround a steady region in the flow with very low TKE, are extremely thin and with low intensity. This supports the suggestion of small scale billowing at the boundaries of the steady region into the surrounding freestream, which is showing similar behaviour to that found by \cite{Chang-et-Constantinescu-2015}. The Sierpinski based obstacles ($d-f$) also display interesting behaviour, with the immediate downstream steady wake containing very low TKE, showing a steady region with negligible turbulent structures, which is initially penetrated by low $K^*$ from the jets. Once the steady region ends it is then followed by much higher TKE as the flow reattaches to create a vortex street.\\

\subsection{Power Spectral Density of Wake}

The Power Spectral Density (PSD) of the wake of each obstacle was calculated using the ``pwelch" method in \textit{Matlab}, whilst applying the ``hammering" function to ensure an even spread of weighting across the whole window. The window size was set at 2000 data points (20 s) with the overlap between the windows being set at 1000 data points. The number of discrete Fourier transform points (NFFT) was set to 4028 to ensure sufficient resolution and the sampling frequency was 100 Hz. The spectral variance was checked with the expected energy from the area below figure to ensure that any discrepancy in the spectra was kept at 1$\pm$0.1. 

\begin{figure}[H]
\begin{subfigure}[First Iteration]{
\includegraphics[width=0.3\textwidth]{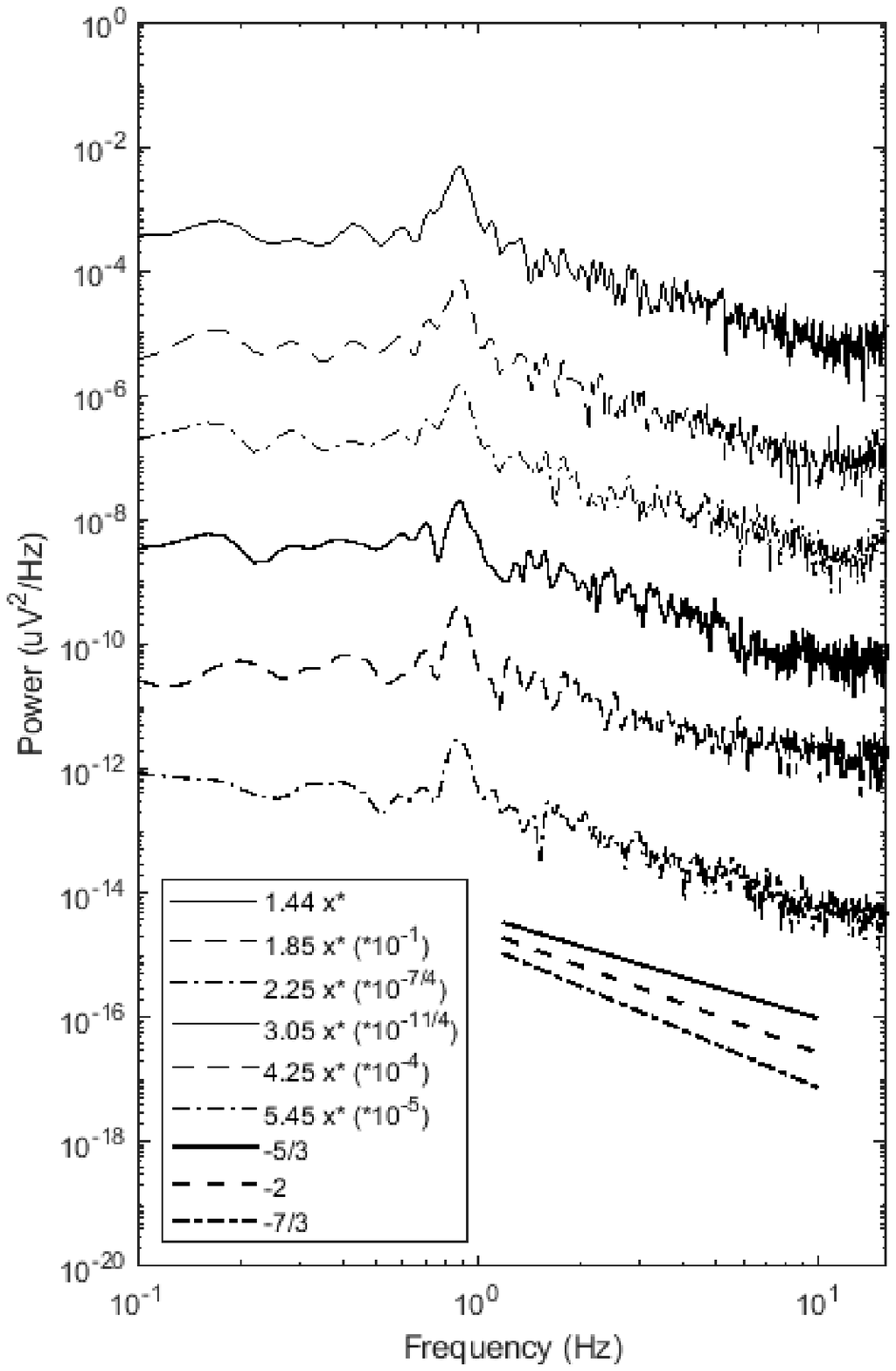}
}
\end{subfigure}
\begin{subfigure}[Solid Column]{
\includegraphics[width=0.31\textwidth]{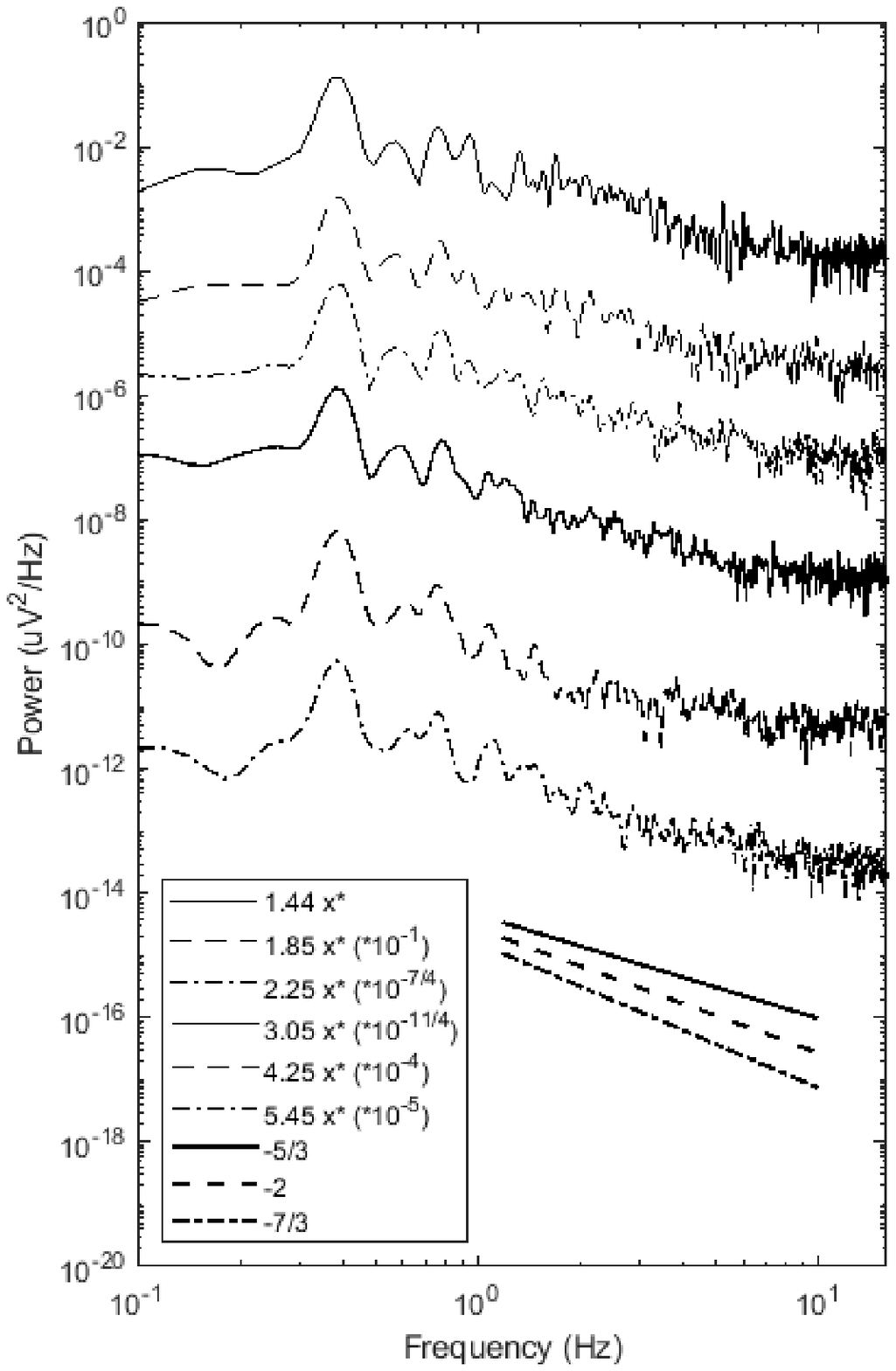}
}
\end{subfigure}
\begin{subfigure}[Uniform Obstacle]{
\includegraphics[width=0.3\textwidth]{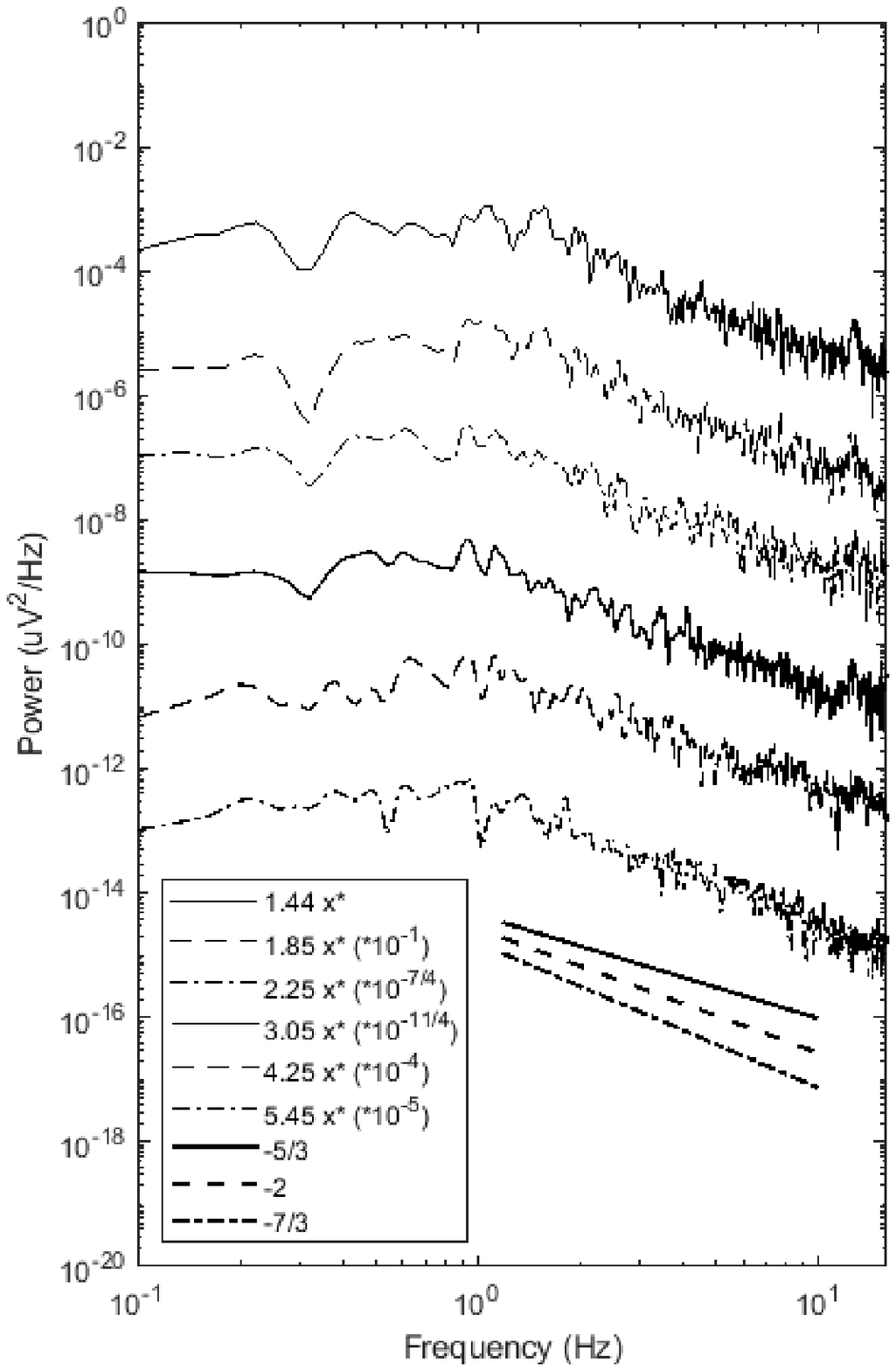}
}
\end{subfigure}
\begin{subfigure}[Sierpinski]{
\includegraphics[width=0.3\textwidth]{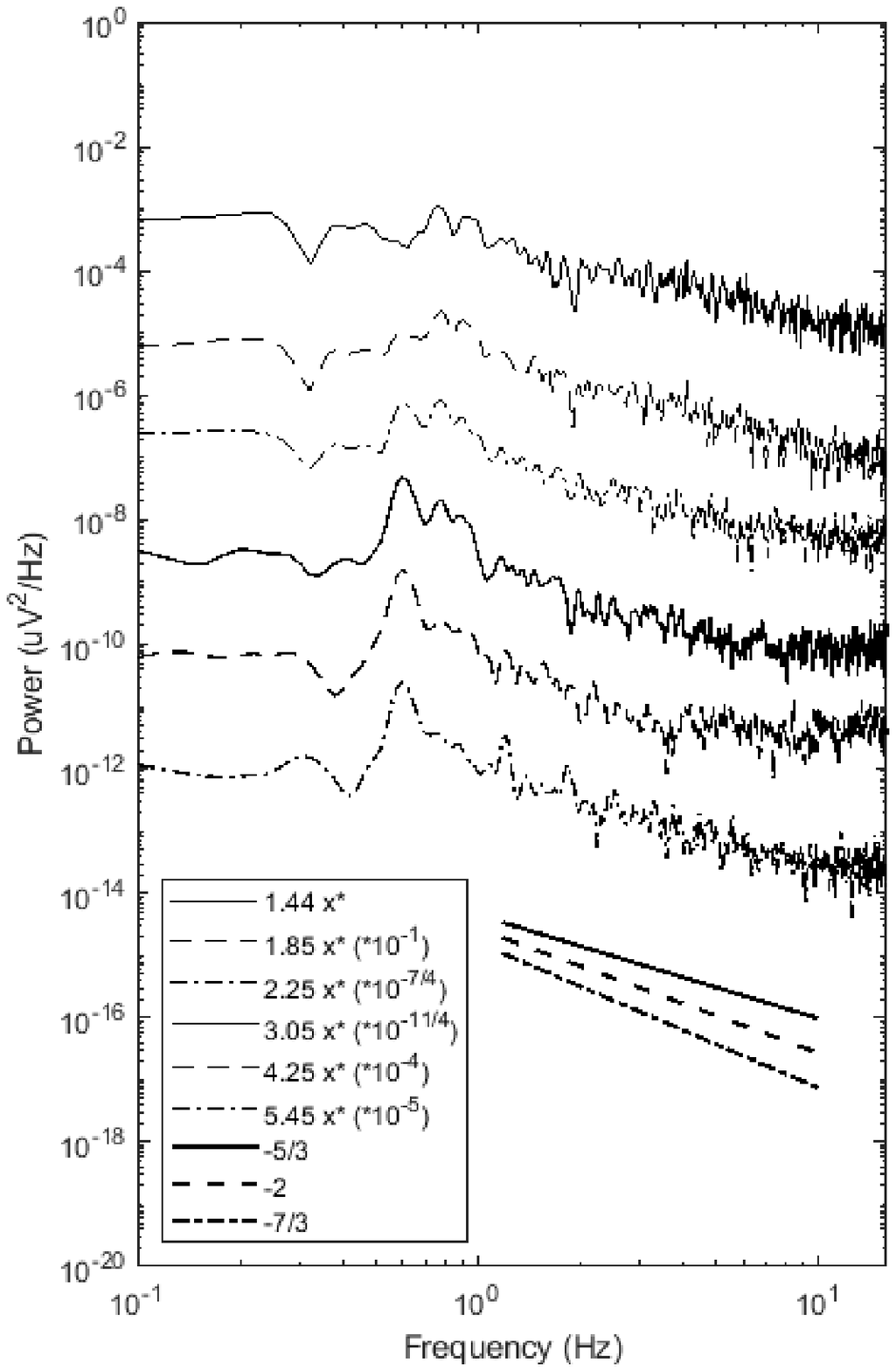}
}
\end{subfigure}
\begin{subfigure}[Upstream Obstacle]{
\includegraphics[width=0.31\textwidth]{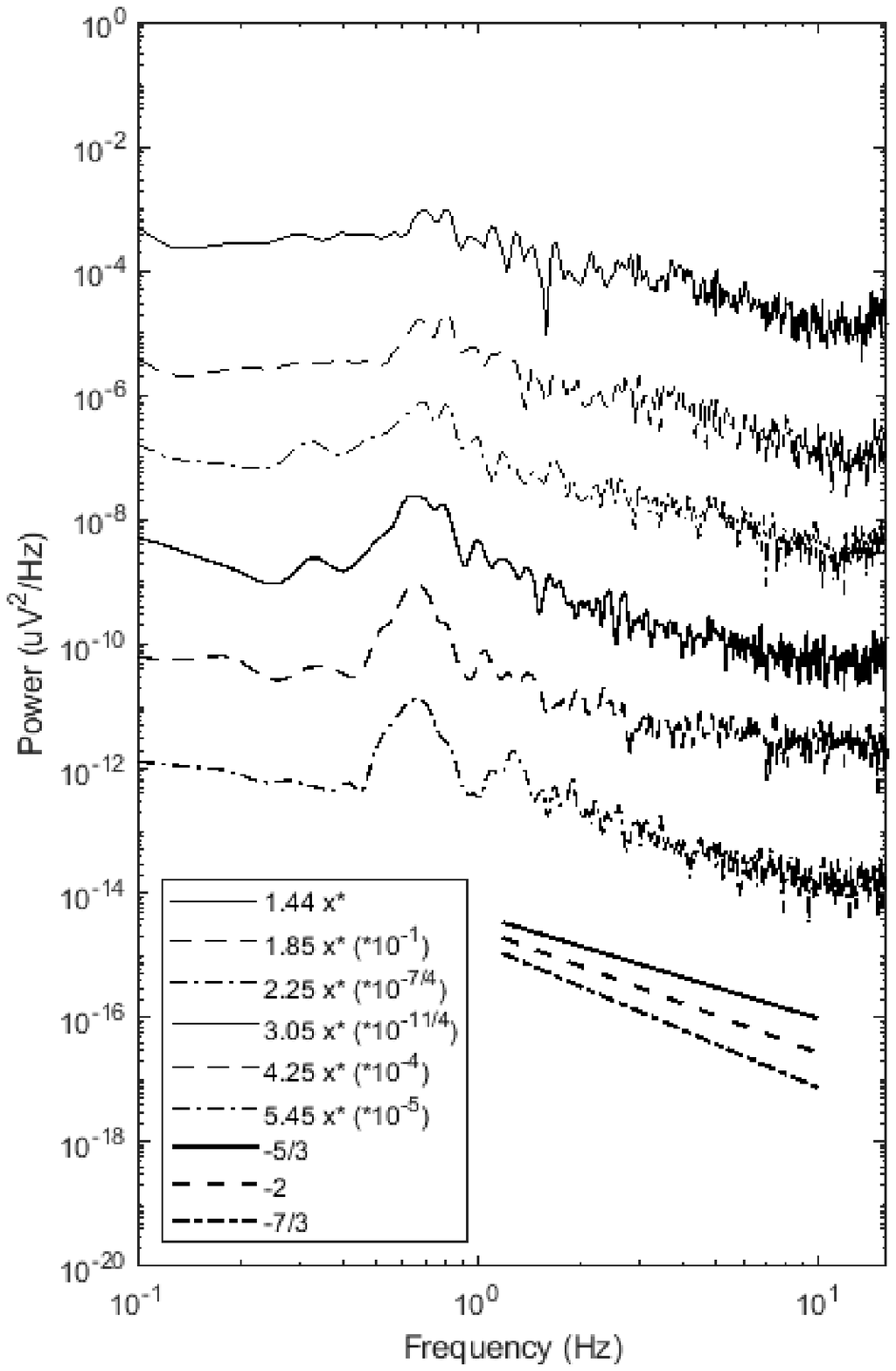}
}
\end{subfigure}
\begin{subfigure}[Downstream Obstacle]{
\includegraphics[width=0.31\textwidth]{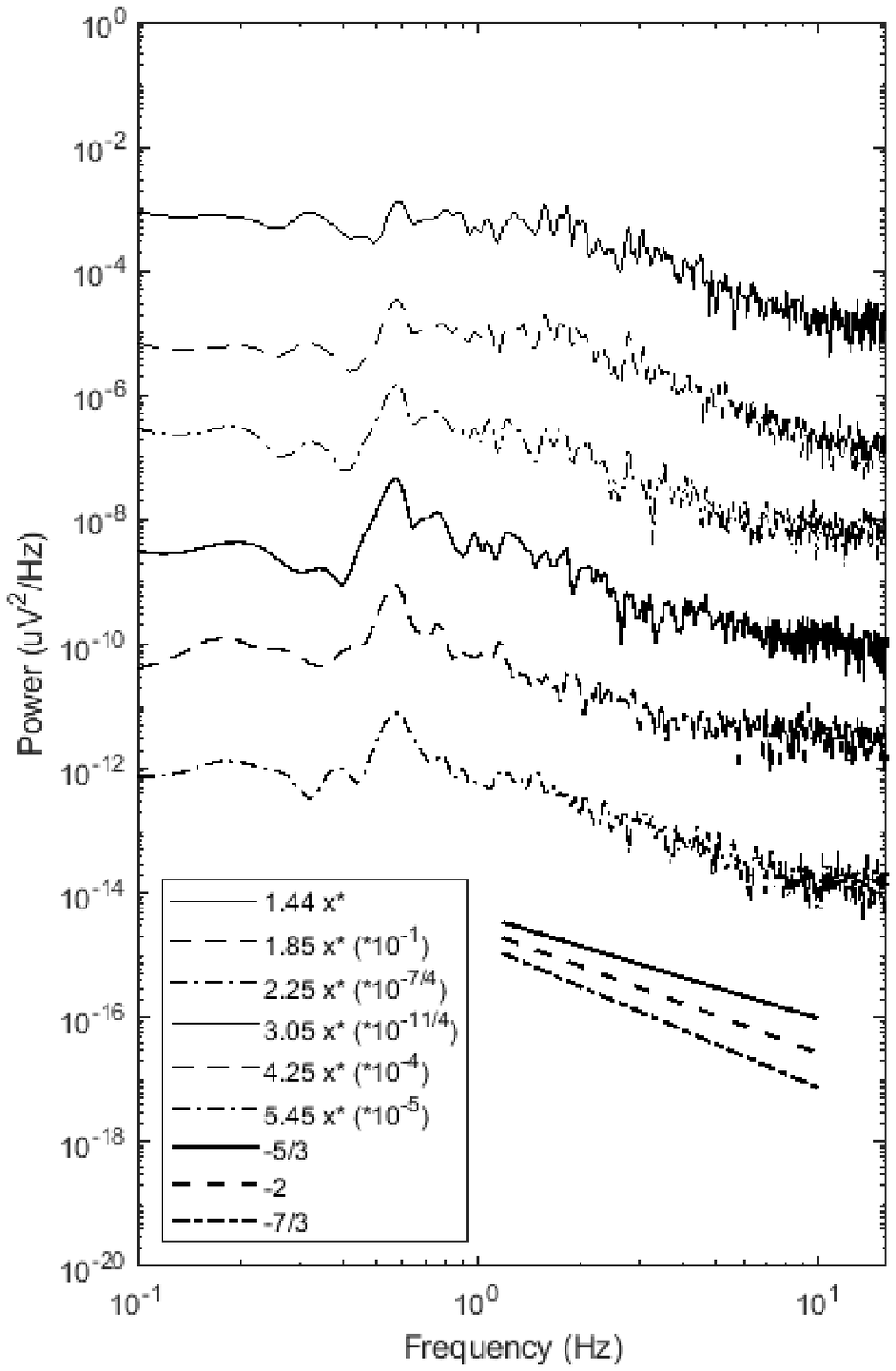}
}
\end{subfigure}
\caption{\label{spectrafig}Power spectral density of all the obstacles tested at varying positions downstream.}
\end{figure}

The resulting v-spectra's can be seen in  Figure~\ref{spectrafig}.
The PSD was calculated at a number of points along the centreline in the downstream wake of each obstacle, allowing the comparison of the turbulent decay in the near field. The solid line represents a gradient of -5/3 which was added to the figures to represent Kolmogorov's power law. This was expected to be seen in the inertial range of the spectra, which for these obstacles was between 1-8~Hz, dependent upon the obstacle.

Using these spectra allows the identification of the dominant frequency of vortex shedding in the flow. This can be seen clearly in each of the plots as the first major peak (moving from left to right). This is difficult to identify in the uniform obstacle ($c$) because of the protracted nature of the steady wake, and lack of reattachment in the near field.

Each of the obstacles displayed their own characteristic spectra reflecting their particular wake. The solid obstacles have pronounced vortex shedding peaks at 0.39 Hz for the solid column ($b$) and 0.88 Hz for the first iteration ($a$). The first iteration obstacle ($a$) displayed a sharper peak suggesting power distribution in a more concentrated range, whereas the larger solid obstacle ($b$) has a broader peak. These two obstacles followed Kolmogorov's power law as expected. 

As previously discussed, the spectra for the uniform obstacle (c) had little to reveal because of the steady region extending the length of the near field. With the dominant power in the spectra being lower than the comparable obstacles at approximately $10^{-3}$.

The Sierpinski based obstacles ($d-f$) displayed interesting power spectra, which can be seen in Figure~\ref{spectrafig}d-f. Each of these obstacles display a dominant vortex shedding frequency of 0.58 Hz for the upstream obstacle ($e$), 0.6 Hz for the Sierpinski obstacle ($d$) and 0.66 Hz for the downstream obstacle ($f$). The upstream obstacle ($e$) can also be seen to exhibit a broader peak which can been associated with the position of the largest iteration being further downstream in the obstacle and the less dominant effect of the obstacles external scale ($D$). 

%

\begin{figure}[H]
%
%
\begin{tikzpicture}

\begin{axis}[%
width=5in,
height=1.5in,
at={(0.758in,0.481in)},
scale only axis,
xmin=0,
xmax=7,
xtick={1,2,3,4,5,6,7},
xticklabels={{(b)Solid},{(c)Uniform},{(e)Up},{(d)Sierpinski},{(f)Down},{(a)First Iter},{},{}},
ymin=0,
ymax=1,
ylabel style={font=\color{white!15!black}},
ylabel={PSD Peak Frequency (Hz)},
axis background/.style={fill=white}
]
\addplot [color=black, mark=square*, mark options={solid, fill=black, black}, forget plot]
  table[row sep=crcr]{%
1	0.39\\
2	0.42\\
3	0.58\\
4	0.6\\
5	0.66\\
6	0.88\\
};
\end{axis}
\end{tikzpicture}%
\caption{Peak frequency of all obstacles in the experiment taken from PSD}
\end{figure}
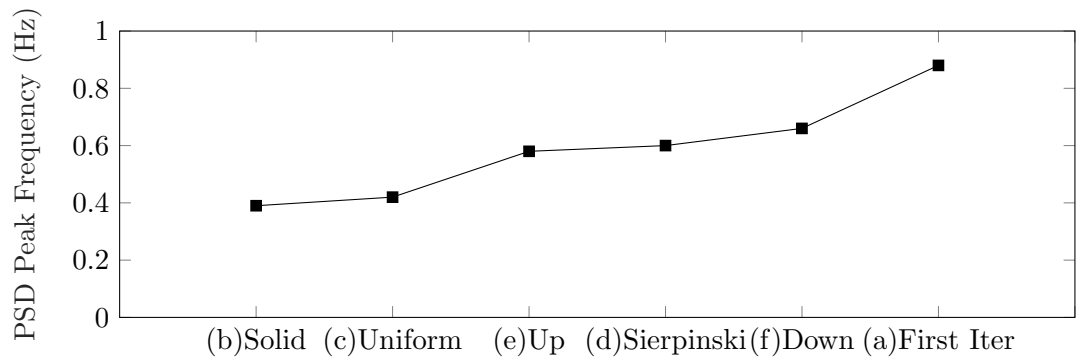

When these dominant frequency values from the PSD are plotted an interesting pattern can be seen in figure 13. The 3DMPOs lie between the solid column and first iteration. But if we look closer at the Sierpinski based obstacles ($d-f$), as the first iteration ($a$) in the obstacle moves downstream, within the fractal obstacle, it becomes more dominant in the peak frequency - with the peak frequency in ($d-f$) moving to meet the first iteration ($a$).

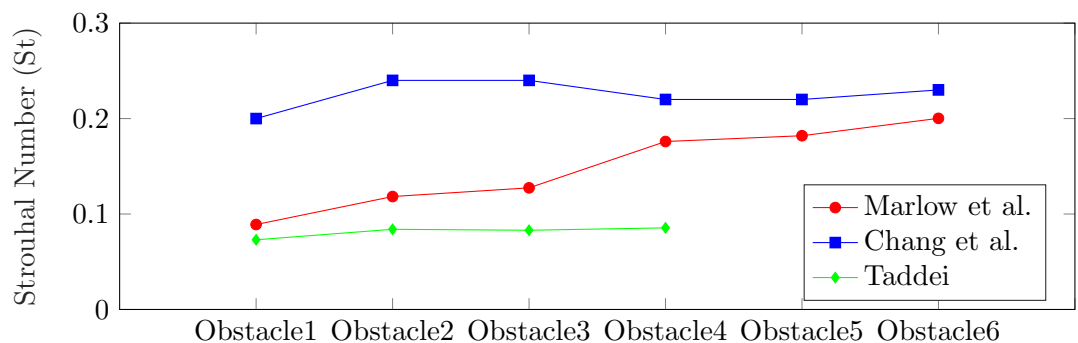
\begin{figure}[H]
%
%
\begin{tikzpicture}

\begin{axis}[%
width=5in,
height=1.5in,
at={(0.758in,0.481in)},
scale only axis,
xmin=0,
xmax=7,
xtick={1,2,3,4,5,6,7},
xticklabels={{Obstacle1},{Obstacle2},{Obstacle3},{Obstacle4},{Obstacle5},{Obstacle6},{},{}},
ymin=0,
ymax=0.3,
ylabel style={font=\color{white!15!black}},
ylabel={Strouhal Number (St)},
axis background/.style={fill=white},
legend style={at={(0.97,0.03)}, anchor=south east, legend cell align=left, align=left, draw=white!15!black}
]
\addplot [color=red, mark=*, mark options={solid, fill=red, red}]
  table[row sep=crcr]{%
1	0.0889887640449438\\
2	0.118314606741573\\
3	0.127415730337079\\
4	0.175955056179775\\
5	0.182022471910112\\
6	0.200224719101124\\
};
\addlegendentry{Marlow et al.}

\addplot [color=blue, mark=square*, mark options={solid, fill=blue, blue}]
  table[row sep=crcr]{%
1	0.2\\
2	0.24\\
3	0.24\\
4	0.22\\
5	0.22\\
6	0.23\\
};
\addlegendentry{Chang et al.}

\addplot [color=green, mark=diamond*, mark options={solid, fill=green, green}]
  table[row sep=crcr]{%
1	0.073\\
2	0.084\\
3	0.083\\
4	0.0855\\
};
\addlegendentry{Taddei}

\end{axis}
\end{tikzpicture}%
\caption{\label{Stouhalfig}Strouhal number comparison of experiment to literature as follows. Marlow et al. (3DMPOs $\phi$ = 0.7, $\Lambda$ ): -, -, 0.08, 0.14, 1.22, 1.22 . Chang et al.(SVF):1, 0.2, 0.2(2d), 0.1,0.05, 0.023. Taddei ($\phi$): 0.05, 0.1, 0.16, 0.24.  }
\end{figure}

From the power spectral density, the Strouhal number $St$ was calculated using the main shedding frequency identified ($f_{peak}$) in Equation~\ref{Stouhaldefeq}. Results are presented in Figure~\ref{Stouhalfig}.

The uniform obstacle's ($c$) Strouhal number was calculated from its dominant frequency $f_{peak}=0.42$~Hz giving a $St$ of 0.13 which is comparable to the $St$ of 0.2 found by simulation of a flow through a cylindrical uniform patch by \cite{Chang-et-Constantinescu-2015}. It is interesting to see that the position of the largest iteration effects the associated $St$. Moving the largest iteration downstream increases $St$, which is the opposite of what is expected. It would be predicted that as the largest iteration is moved downstream it would have a more dominating affect on the shedding frequency and as a result, move the $St$ of the Sierpinski based obstacle towards the first iteration $St$. It, in fact, moves the Sierpinski based obstacles $St$ further away from the comparable first iteration $St$.

\section{Conclusion \label{seccon}}

We compared the wake structure of 3 cases: (1) non-porous, (2) porous obstacles with a single internal scale and (3) porous fractals.
Each porous obstacle had the same void fraction ($\simeq 0.7$) but a different distribution of internal scales.

\begin{figure}[H]
\includegraphics[angle=90,scale=0.32]{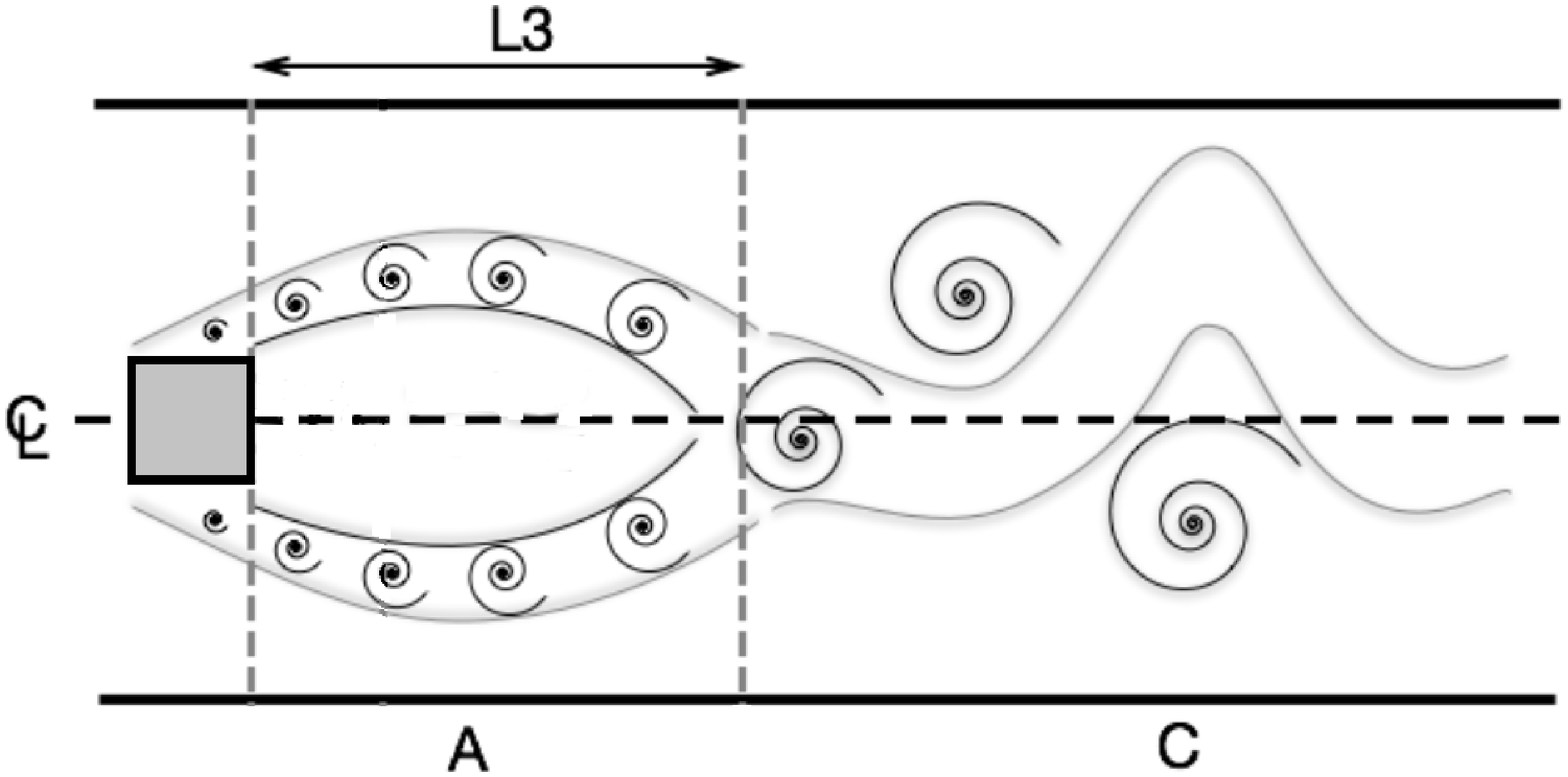}
\includegraphics[angle=90,scale=0.64]{uniformschematic}
\includegraphics[angle=90,scale=0.64]{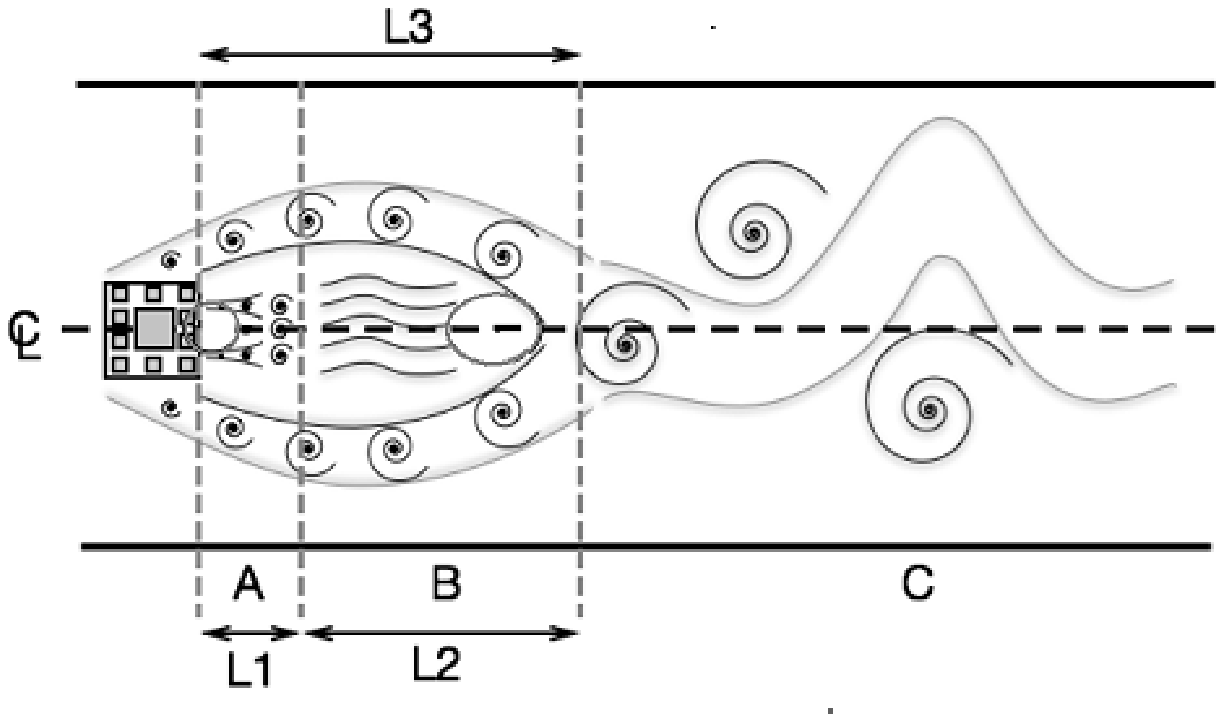}
\\
\hspace*{0.1\textwidth} a)   
\hspace{0.25\textwidth} b) \hspace{0.3\textwidth} c) 
\caption{\label{wakepatsum3fig}Illustrated wake features behind the obstacles tested.}
\end{figure}

The PIV results show that there are similarities between each of the obstacles in terms of the structures formed behind the obstacle but the internal scale distribution is responsible for determining the particular wake characteristics. Therefore the void fraction is not enough as a single parameter to characterise the wake structure. As summarised in Figure~\ref{wakepatsum3fig}, the obstacles internal scales lead to the formation of unique wake patterns 
depending on the respective $D_f$, $\Lambda$ and $\sigma$ of the 3DMPOs.

The general behaviour involves lateral bleeding from the obstacles leading to the formation of two separated shear layers at the obstacle lateral boundaries. The lateral bleeding reduces the stream-wise velocity by around $0.2u^*$ causing the downstream extension of low-velocity legs which encase a medium-velocity ($0.5u^*$) steady region. 


As shown by the turbulent kinetic energy maps (Figure~\ref{TKEfig}) there is very negligible turbulent behaviour in this so-called steady region, which is only penetrated by low magnitude turbulence resulting from the streamwise jets. The lateral bleeding in the uniform obstacle causes extended low-velocity legs that are not seen to reattach in the near field. These low-velocity legs then cause small scale billowing into the surrounding flow (see Figure~\ref{wakepatsum3fig} b). 

The significantly longer steady region observed behind the uniform obstacle is thought to be caused by the smaller $D_f$ \& $\Lambda$ representing the uniform nature of gaps that create internal corridors channelling the flow downstream. These corridors lead to the formation of high-velocity stream-wise jets that extend the steady region indefinitely downstream. The jets are a product of the internal scales associated with the obstacle, with case (2) $K^*_{max}$ = 0.03 \& case (3) $K^*_{max}$ = 0.2. This indicates the beneficial nature of using fractals in mixing applications, however low $D_f$ \& $\Lambda$ are to be preferred when looking to reduce the downstream effect of the obstacle.


The Sierpinski based obstacles observe a steady core region immediately downstream of the obstacle caused by the high-velocity streamwise jets (see Figure~\ref{wakepatsum3fig} c). These jets push the typical recirculation region as seen in the solid cases downstream. It can also be seen that the largest iteration in the Sierpinski based obstacles is also responsible for the creation of its own recirculation region within the steady wake. The position of the largest internal scale and the resultant $\sigma$ of the fractal obstacles seems to have a driving factor on the downstream wake characteristics, determining the $L_3$ in fractal obstacles, with Upstream ($\sigma$ = 0.763) $L_3$ = 4, Sierpinski ($\sigma$ = 0.702) $L_3$ = 4.5  and Downstream ($\sigma$ = 0.642) $L_3$ = 5.

Once the two separated shear layers have reattached a Reynolds shear stress increase indicates the formation of a vortex street as typically observed behind a solid obstacle. The frequency $f$ of this oscillating vortex street was slightly affected by the position of the largest iteration and $\sigma$ in the obstacle, with the upstream obstacle reducing the oscillation frequency, and the downstream obstacle increasing it when compared to the Sierpinski obstacle as we have seen in figure 13. It can be seen that the largest scale iteration in the Sierpinski plays a significant role in the determination of the $f$. However, spectral analysis shows that this vortex street is at a lower intensity ($R_{uv}^*=0.04$) than that observed behind the solid obstacles ($R_{uv}^*=0.1$).
\\[2ex]
The porous obstacles were seen to renounce the traditional Kolmogorov power law for the energy spectra.
The energy cascade in the inertial region seemed to follow larger gradients of either -2 or -7/3. The ability of these obstacles to distribute energy between the spatial modes as well as transfer energy down the scales at an increased rate has numerous advantages when it comes to flow control and mixing. It can allow the turbulent frequencies and structures formed to be controlled and designed to minimise the effect of the downstream wake. 

%
%
%

\bibliographystyle{wsnatbib}

\end{document}